# Multiscale modeling of stress transfer in hybrid composites

**S. I. Kundalwal[a,*], S. Kumar [a,b,#]**

[a]*Department of Mechanical and Materials Engineering, Masdar Institute of Science and Technology, Abu Dhabi 54224, UAE*
[b]*Department of Mechanical Engineering, Massachusetts Institute of Technology, Cambridge, MA 02139-4307, USA*

## ABSTRACT

This study is focused on the mechanical properties and stress transfer behavior of multiscale composites containing nano- and micro-scale reinforcements. The distinctive feature of construction of this composite is such that the carbon nanostructures (CNS) are dispersed in the matrix around the continuous microscale fiber to modify microfiber-matrix interfacial adhesion. Such CNS are considered to be made of aligned CNTs (A-CNTs). Accordingly, multiscale models are developed for such hybrid composites. First, molecular dynamics simulations in conjunction with the Mori-Tanaka method are used to determine the effective elastic properties of nano-engineered interphase layer composed of CNS and epoxy. Subsequently, a micromechanical pull-out model for a continuous fiber multi-scale composite is developed, and stress transfer behavior is studied for different orientations of CNS considering their perfect and imperfect interfacial bonding conditions with the surrounding epoxy. Such interface condition was modeled using the linear spring layer model with a continuous traction but a displacement jump. The current pull-out model accounts for the radial as well as the axial deformations of different orthotropic constituent phases of the multiscale composite. The results from the developed pull-out model are compared with those of the finite element analyses and are found to be in good agreement. Our results reveal that the stress transfer characteristics of the multiscale composite are significantly improved by controlling the CNT morphology around the fiber, particularly, when they are aligned along the axial direction of the microscale fiber. The results also show that the CNS-epoxy interface weakening significantly influences the radial stress along the length of the microscale fiber.

[*]Banting Fellow at the Mechanics and Aerospace Design Laboratory, Department of Mechanical and Industrial Engineering, University of Toronto, Toronto, Canada

## 1. Introduction

The structural performance of a composite under service load is largely affected by the fiber-matrix interfacial properties. The ability to tailor interfacial properties is essential to ensure efficient load transfer from matrix to the reinforcing fibers, which help to reduce stress concentrations and improve overall mechanical properties of a resulting composite (Zhang et al., 2012). Several experimental and analytical techniques have been developed thus far to gain insights into the basic mechanisms dominating the fiber-matrix interfacial characteristics. The strength and toughness of the resulting composite is dependent on two facts: (i) efficient stress transfer from matrix to fiber, and (ii) the nature of fiber-matrix interface. To characterize these issues, the pull-out test or shear lag model is typically employed. A number of analytical and computational two- and three-cylinder pull-out models have been developed to better understand the stress transfer mechanisms across the fiber-matrix interface (Kim et al., 1992; Kim et al., 1994; Tsai and Kim, 1996; Quek and Yue, 1997; Fu et al., 2000; Fu and Lauke, 2000; Banholzer et al., 2005; Ahmed and Keng, 2012; Meng and Wang, 2015; Upadhyaya and Kumar, 2015). These models differ in terms of whether the interphase between the fiber and the matrix is considered or not, and whether we are concerned with long or short fiber composites. In the case of three-cylinder pull-out model, a thin layer of interphase, formed as a result of physical and chemical interactions between the fiber and the matrix, is considered. The chemical composition of such an interphase differs from both the fiber and matrix materials but its mechanical properties lie between those of the fiber and the matrix (Drzal, 1986; Sottos et al., 1992; Kundalwal and Meguid, 2015), and such nanoscale interphase has a marginal influence on the bulk elastic properties of a composite. On the other hand, a relatively thick interphase can be engineered between the fiber and the matrix, especially a third phase made of different material than the main constituent phases (see for e. g. Liljenhjerte and Kumar, 2015). Such microscale interphase strongly influences the mechanical and interfacial properties of a composite, where the apparent reinforcing effect is related to the cooperation of the interfacial adhesion strength, and the interphase serving to inhibit crack propagation or as mechanical damping elements [see Zhang et al. (2010) and the references therein].

Recently, CNTs and graphene have attracted intense research interest because of their remarkable electro-thermo-mechanical properties, which make them candidates as nano-fillers in composite materials (Ray and Kundalwal, 2013a,b; Kundalwal et al., 2014a,b,c; Pal and Kumar, 2016a, b; Cui et al., 2016; Arif et al., 2016; Kumar et al., 2016). Extensive research has been dedicated to the introduction of graphene and CNTs as the modifiers to the conventional composites in order to enhance their multifunctional properties. For example, Bekyarova et al. (2007) reported an approach to the development of advanced structural composites based on engineered CNT-microscale fiber reinforcement; the CNT-carbon fabric-epoxy composites showed ~30% enhancement of the interlaminar shear strength as compared to that of microscale fiber-epoxy composites. Cho et al. (2007) modified the epoxy matrix in microscale fiber-epoxy composites with graphite nanoplatelets and reported the improved in-plane shear properties and compressive strength for the resulting hybrid composite. Wardle and his co-authors (2008; 2012; 2014) grew aligned CNTs (A-CNTs) on the circumferential surfaces of microfibers to reinforce the matrix and reported the improvement in composite delamination resistance, toughness, Mode I fracture toughness, interlaminar shear strength, matrix-dominated elastic properties and electrical conductivity. Hung et al. (2009) fabricated unidirectional composite in which CNTs were directly grown on the circumferential surfaces of conventional microscale fibers. Davis et al. (2010) fabricated the carbon fiber reinforced composite incorporating functionalized CNTs in the epoxy matrix; as a consequence, they observed significant improvements in tensile strength, stiffness and resistance to failure due to cyclic loadings. Zhang et al. (2010) deposited CNTs on the circumferential surfaces of electrically insulated glass fiber surfaces. According to their fragmentation test results, the incorporation of an interphase with a small number of CNTs around the fiber, remarkably improved the interfacial shear strength of the fiber-epoxy composite. The functionalized CNTs were incorporated by Davis et al. (2011) at the fiber/fabric–matrix interfaces of a carbon fiber reinforced epoxy composite laminate material; their study showed improvements in the tensile strength and stiffness, and resistance to tension–tension fatigue damage due to the created CNT reinforced region at the fiber/fabric–matrix interfaces. A numerical method is proposed by Jia et al. (2014) to theoretically investigate the pull-out of a hybrid fiber coated with CNTs. They developed two-step finite element (FE) approach: a single CNT pull-out from the matrix at microscale and the pull-out of the hybrid fiber at macroscale. Their numerical results indicate that the apparent interfacial shear strength of the hybrid fiber and the specific pull-out energy are significantly increased due to the additional bonding of the CNT–matrix interface. A beneficial interfacial effect of the presence of CNTs on the circumferential surface of the microscale fiber samples is demonstrated by Jin et al. (2014) resulting in an increase in the maximum interlaminar shear strength (>30 MPa) compared to uncoated samples.

This increase is attributed to an enhanced contact between the resin and the fibers due to an increased surface area as a result of the CNTs. To improve the interfacial properties of microscale fiber-epoxy composites, Chen et al. (2015) introduced a gradient interphase reinforced by graphene sheets between microscale fibers and matrix using a liquid phase deposition strategy; due to the formation of this gradient interphase, 28.3% enhancement in interlaminar shear strength of unidirectional microscale fiber-epoxy composites is observed with 1 wt,% loading of graphene sheets. Recently, two types of morphologies are investigated by Romanov et al. (2015): CNTs grown on fibers and CNTs deposited in fiber coatings. The difference in the two cases is the orientation of CNTs near the fiber interface: radial for grown CNTs and tangent for CNTs in the coatings.

Findings in the literature indicate that the use of nano-fillers and conventional microscale fibers together, as multiscale reinforcements, significantly improve the overall properties of multiscale composites, which are unachievable in conventional composites. As is well known, damage initiation is progressive with the applied load and that the small crack at the fiber-matrix interface may reduce the fatigue life of composites. By toughening the interfacial fiber-matrix region with nano-fillers, we can increase the damage initiation threshold and long-term reliability of conventional composites. This concept can be utilized to grade the matrix properties around the microscale fiber, which may eventually improve the stress transfer behaviour of multiscale composite. To the best of our knowledge, there has been no pull-out model to study the stress transfer characteristics of multiscale composite containing transversely isotropic nano- and micro-scale fillers. This is indeed the motivation behind the current study. The current study is devoted to the development of a pull-out model for analyzing the stress transfer characteristics of multiscale composite. A-CNT bundles reinforced in the polymer (epoxy thermoset) material is considered as a special case of carbon nanostructures (CNS) embedded between the fiber and matrix, most relevant to bundling of single-wall carbon nanotubes (SWCNTs); the resulting intermediate phase, containing CNS and epoxy, is considered as an interphase. First, we carried out multiscale study to determine the transversely isotropic elastic properties of an interphase through MD simulations in conjunction with the Mori-Tanaka model. Then the determined elastic moduli of the interphase are used in the development of three-phase pull-out model. Particular attention is paid to investigate the effect of orientations of CNS considering their perfect and imperfect interfacial bonding conditions with the surrounding epoxy on the stress transfer characteristics of multiscale composite.

## 2. Multiscale modeling

For most multiscale composites, mechanical response and fracture behavior arise from the properties of the individual constituents at each level as well as from the interaction between these constituents across different length scales. As a consequence, different multiscale modeling techniques have been developed over the last decade to predict the continuum properties of composites at the microscale (Tsai et al., 2010; Yang et al., 2012; Alian et al., 2015a,b). Here, multiscale modeling of CNS-reinforced epoxy interphase is achieved in two consecutive steps: (i) elastic properties of the CNS comprised of a bundle of CNTs and epoxy molecules are evaluated using molecular dynamics (MD) simulations; (ii) the Mori-Tanaka method is then used to calculate the bulk effective properties of the nano-engineered interphase layer.

### *2.1 Molecular modeling*

This section describes the procedure for building a series of MD models for the epoxy and the CNS. The technique for creating an epoxy and CNS is described first, followed by the MD simulations for determining the isotropic elastic properties of the epoxy material and the transversely isotropic elastic properties of the CNS. All MD simulations runs are conducted with large-scale atomic/molecular massively parallel simulator (LAMMPS; Plimpton, 1995). The consistent valence force field (CVFF; Dauber-Osguthorpe, 1998) is used to describe the atomic interactions between different atoms. The CVFF has been used by several researchers to model the CNTs and their composite systems (Tunvir et al., 2008; Li et al., 2012; Kumar et al., 2014). A very efficient conjugate gradient algorithm is used to minimize the strain energy as a function of the displacement of the MD systems while the velocity Verlet algorithm is used to integrate the equations of motion in all MD simulations. Periodic boundary conditions have been applied to the MD unit cell faces. Determination of the elastic properties of the pure epoxy and the CNS is accomplished by straining the MD unit cells followed by constant-strain energy minimization. The averaged stress tensor of the MD unit cell is defined in the form of virial stress (Allen and Tildesley, 1987); as follows

$$\sigma = \frac{1}{V}\sum_{i=1}^{N}\left(\frac{m_i}{2}{v_i}^2 + F_i\, r_i\right) \tag{1}$$

where V is the volume of the unit cell; $v_i$, $m_i$, $r_i$ and $F_i$ are the velocity, mass, position and force of the $i^{th}$ atom, respectively.

### *2.1.1 Modeling of EPON 862-DETDA epoxy*

Thermosetting polymers are the matrices of choice for structural composites due to their high stiffness, strength, creep resistance and thermal resistance when compared with thermoplastic polymers (Pascault et al., 2002). These desirable properties stem from the three-dimensional (3D) crosslinked structures of these polymers. Many thermosetting polymers are formed by mixing a resin (epoxy, vinyl ester, or polyester) and a curing agent. We used epoxy material based on EPON 862 resin and Diethylene Toluene Diamine (DETDA) curing agent to form a crosslinked structure, which is typically used in the aerospace industry. The molecular structures of these two monomers are shown in Fig. 1. To simulate the crosslinking process, the potential reactive sites in the epoxy resin can be activated by hydrating the epoxy oxygen atoms at the ends of the molecule, see Fig. 1(b). The EPON 862-DETDA weight ratio was set to 2:1 to obtain the best elastic properties (Bandyopadhyay and Odegard, 2012). The initial MD unit cell, consisting of both activated epoxy (100 molecules of EPON 862) and a curing agent (50 molecules of DETDA), was built using the PACKMOL software (Martínez et al., 2009). The polymerization process usually occurs in two main stages: pre-curing equilibration and curing of the polymer network. The main steps involved in determining the elastic moduli of pure epoxy are described as follows:

**Step 1 (pre-curing equilibration):** The initial MD unit cell was compressed gradually through several steps from its initial size of 50 Å × 50 Å × 50 Å to the targeted dimensions of 39 Å × 39 Å × 39 Å. The details of the MD unit cell are summarized in Table 1. At each stage, the atoms coordinates are remapped to fit inside the compressed box then a minimization simulation was performed to relax the coordinates of the atoms. The system was considered to be optimized once the change in the total potential energy of the system between subsequent steps is less than $1.0\times10^{-10}$ kcal/mol (Alian et al., 2015b). The optimized system is then equilibrated at room temperature in the constant temperature and volume canonical (NVT) ensemble over 100 ps by using a time step of 1 fs.

**Step 2 (curing):** After the MD unit cell is fully equilibrated in step 1, the polymerization and crosslinking is simulated by allowing chemical reactions between reactive atoms. Chemical reactions are simulated in a stepwise manner using a criterion based on atomic distances and the type of chemical primary or secondary amine reactions as described elsewhere (Bandyopadhyay and Odegard, 2012). The distance between all pairs of reactive C–N atoms are computed and new bonds are created between all those that fall within a pre-assigned cut-off distance. We considered this distance to be 5.64 Å, four times the equilibrium C–N bond length (Varshney et al., 2008; Li and Strachan, 2010). After the new bonds are identified, all new additional covalent terms were created and hydrogen atoms from the reactive C and N atoms were removed. Then several 50 ps isothermal–isobaric (NPT) simulations are preformed until no reactive pairs are found within the cut-off distance. At the end, the structure is again equilibrated for 200 ps in the NVT ensemble at 300 K.

**Step 3 (elastic coefficients):** After the energy minimization process, the simulation box was volumetrically strained in both tension and compression to determine the bulk modulus by applying equal strains in the loading directions along all three axes; while the average shear modulus was determined by applying equal shear strains on the simulation box in xy, xz, and yz planes. In all simulations, strain increments of 0.25% have been applied along a particular direction by uniformly deforming or shearing the simulation box and updating the atom coordinates to fit within the new dimensions. After each strain increment, the MD unit cell was equilibrated using the NVT ensemble at 300 K for 10 ps. It may be noted that the fluctuations in the temperature and potential energy profiles are less than 1% when the system reached equilibrium after about 5 ps (Haghighatpanah and Bolton, 2013) and several existing MD studies (Frankland et al., 2003; Tsai et al., 2010; Haghighatpanah and Bolton, 2013; Alian et al., 2015b) used 2 ps to 10 ps time step in their MD simulations to equilibrate the systems after each strain increment. Then, the stress tensor is averaged over an interval of 10 ps to reduce the effect of fluctuations. These steps have been repeated again in the subsequent strain increments until the total strain reached up to 2.5%. Based on the calculated bulk and shear moduli, Young's modulus (E) and Poisson's ratio (ν) are determined. The predicted elastic properties of the epoxy using MD simulations are summarized in Table 2 and Young's modulus is found to be consistent with the experimentally measured modulus of a similar epoxy (Morris, 2008).

### *2.1.2 MD simulations of CNS*

Despite the great potential of applying CNTs in composite materials, an intrinsic limitation in directly scaling up the remarkable elastic properties of CNTs is due to their poor dispersion, agglomeration and aggregation (Barai and Weng, 2011). It is difficult to uniformly disperse CNTs in the matrix during the fabrication process and the situation becomes more challenging at high CNT loadings. This is attributed to the fact that CNTs, particularly SWCNTs, have a tendency to agglomerate and aggregate into bundles due to their high surface energy and surface area (Dumlich et al., 2011). Therefore, we consider the epoxy nanocomposite reinforced with A-CNT bundles, which is more practical and realistic representation of embedded CNTs. The MD unit cell is constructed to represent an epoxy nanocomposite containing a bundle of thirteen CNTs, as shown in Fig. 2. The initial distance between the adjacent CNTs in the bundle considered was 3.4 Å, which is the intertube separation distance in multi-walled CNTs (Odegard et al., 2003; Qi et al., 2005; Tunvir et al., 2008; Dumlich et al., 2011). The noncovalent bonded CNT-epoxy nanocomposite system is considered herein; therefore, the interactions between the atoms of the embedded CNTs and the surrounding epoxy are solely from non-bonded interactions. These non-bonded interactions between the atoms are represented by van der Waals (vdW) interactions and Coulombic forces. The cut-off distance for the non-bonded interaction was set to 14.0 Å (Haghighatpanah and Bolton, 2013). The unit cell is assumed to be transversely isotropic with the 1–axis being the axis of symmetry; therefore, only five independent elastic coefficients are required to define the elastic stiffness tensor. The cylindrical molecular structure of each CNT is treated as an equivalent solid cylindrical fiber (Odegard et al., 2003; Tsai et al., 2010) for determining its volume fraction in the nanocomposite (Frankland et al., 2003),

$$v_{CNT} \cong \frac{\pi\left(R_{CNT}+\frac{h_{vdW}}{2}\right)^2 N_{CNT}L_{CNT}}{V_{CNS}} \tag{2}$$

where $R_{CNT}$ and $L_{CNT}$ denote the respective radius and length of a CNT; $h_{vdW}$ is the vdW equilibrium distance between a CNT and the surrounding epoxy matrix; $N_{CNT}$ is the number of CNTs in the bundle; and $V_{CNS}$ is the volume of the CNS.

The CNS is constructed by randomly placing the crosslinked epoxy structures around the A-CNT bundle. The details of the CNS are summarized in Table 1. Five sets of boundary conditions have been chosen to determine each of the five independent elastic constants such that a single property can be independently determined for each boundary condition. The displacements applied at the boundary of the CNS are summarized in Table 3; in which symbols have usual meaning. To determine the five elastic constants, the CNS is subjected to four different tests: longitudinal tension, transverse tension, in-plane tension, in-plane shear and out of-plane shear. The steps involved in the MD simulations of the CNS are the same as adopted in the case of pure epoxy. The first row of Table 4 presents the computed effective elastic properties of the CNS through MD simulations. These properties of the CNS will be used as the properties of nanoscale fiber in the micromechanical model to determine the effective elastic moduli of the interphase at the microscale level (see Fig. 3).

## *2.2 Effective elastic properties of CNS – engineered interphase*

In this section, the elastic properties of the pure epoxy and the CNS obtained from the MD simulations are used as input in the Mori-Tanaka model in order to determine the effective elastic properties of the interphase. Fig. 3 shows the schematic cross-section of the three-phase multiscale composite. Around the microscale fiber, the interphase is considered to be made of CNS and epoxy. Here, we consider three different cases of interphase, in which: (i) CNS are aligned along the direction of the microscale fiber, (ii) CNS are aligned radially to the axis of the microscale fiber, and (iii) CNS are randomly dispersed. Considering the CNS as a fiber, the Mori-Tanaka model (Mori and Tanaka, 1973) can be utilized to estimate the effective elastic properties of the interphase, as follows (Benveniste, 1987):

$$[\mathrm{C}] = [\mathrm{C^m}] + \mathrm{v_{CNS}}([\mathrm{C^{CNS}}] - [\mathrm{C^m}])\left([\tilde{\mathrm{A}}_1]\left[\mathrm{v_m}[\mathrm{I}] + \mathrm{v_{CNS}}[\tilde{\mathrm{A}}_1]\right]^{-1}\right) \tag{3}$$

in which

$$[\tilde{\mathrm{A}}_1] = \left[[\mathrm{I}] + [\mathrm{S^E}]([\mathrm{C^m}])^{-1}([\mathrm{C^{CNS}}] - [\mathrm{C^m}])\right]^{-1}$$

where $[\mathrm{C^{CNS}}]$ and $[\mathrm{C^m}]$ are the stiffness tensors of the CNS and the epoxy matrix, respectively; $[\mathrm{I}]$ is an identity tensor; $\mathrm{v_{CNS}}$ and $\mathrm{v_m}$ represent the volume fractions of the CNS and the epoxy matrix, respectively; and $[\mathrm{S^E}]$ indicates the Eshelby tensor. The specific forms of the Eshelby tensor, for the first two cases, are explicitly given in Appendix A.

In the case (iii) of epoxy reinforced with randomly oriented CNS in the three-dimensional space, the following relation can be used to determine the effective elastic properties of the interphase:

$$[\mathrm{C}] = [\mathrm{C^m}] + \mathrm{v_{CNS}}([\mathrm{C^{CNS}}] - [\mathrm{C^m}])\left([\widetilde{\mathrm{A}}_2]\left[\mathrm{v_m}[\mathrm{I}] + \mathrm{v_{CNS}}[\langle[\widetilde{\mathrm{A}}_2]\rangle]\right]^{-1}\right) \tag{4}$$

in which

$$[\widetilde{\mathrm{A}}_2] = \left[[\mathrm{I}] + [\mathrm{S^E}]([\mathrm{C^m}])^{-1}([\mathrm{C^{CNS}}] - [\mathrm{C^m}])\right]^{-1}$$

The terms enclosed with angle brackets in Eq. (4) represent the average value of the term over all orientations defined by transformation from the local coordinate system of the CNS to the global coordinate system. The transformed mechanical strain concentration tensor for the CNS with respect to the global coordinates is given by

$$[\widetilde{\mathrm{A}}_{\mathrm{ijkl}}] = \mathrm{t_{ip}t_{jq}t_{kr}t_{ls}}[\mathrm{A_{pqrs}}] \tag{5}$$

where $\mathrm{t_{ij}}$ are the direction cosines for the transformation and are given in Appendix A.

Finally, the random orientation average of the dilute mechanical strain concentration tensor $\langle[\widetilde{\mathrm{A}}_2]\rangle$ can be determined by using the following relation (Marzari and Ferrari, 1992):

$$\langle[\widetilde{\mathrm{A}}_2]\rangle = \frac{\int_{-\pi}^{\pi}\int_{0}^{\pi}\int_{0}^{\pi/2}[\widetilde{\mathrm{A}}](\phi,\gamma,\psi)\,\sin\gamma\,d\phi d\gamma d\psi}{\int_{-\pi}^{\pi}\int_{0}^{\pi}\int_{0}^{\pi/2}\sin\gamma\,d\phi d\gamma d\psi} \tag{6}$$

where $\phi$, $\gamma$, and $\psi$ are the Euler angles with respect to 1, 2, and 3 axes..

### *2.2.1 The CNS-epoxy interface effect*

We now consider the effect of CNS-epoxy interface on the effective elastic properties of interphase. Equation (4) was derived assuming that the interface between CNS and epoxy is elastically strong and perfect; but in reality, the nano-fiber/matrix interface is imperfect and the effect such an interface is inevitable (Esteva and Spanos, 2009; Barai and Weng, 2011; Pan et al., 2013; Wang et al., 2014). This imperfect condition can have a profound influence on the effectiveness of CNS reinforcement. In order to address this issue in the context of elastic response, Mori-Tanaka model with the modified Eshelby tensor (Qu, 1993) is presented herein. In conjunction with generic composite materials mechanics, Qu (1993) introduced imperfection in the interface using a spring layer of insignificant thickness and finite stiffness. Such layer produces continuous interfacial tractions but discontinuous displacements. The equations that model the interfacial traction continuity and the displacement jump ($\Delta u_i$) at the CNS-epoxy interface can be written as follows:

$$\Delta\sigma_{ij}n_j \equiv \left[\sigma_{ij}(S^+) - \sigma_{ij}(S^-)\right]n_j = 0 \tag{7}$$

$$\Delta u_i \equiv [u_i(S^+) - u_i(S^-)]n_{ij}\sigma_{jk}n_k \tag{8}$$

where S and n represent the CNS-matrix interface and its unit outward normal vector, respectively. The terms $u_i(S^+)$ and $u_i(S^-)$ are the values of the displacements when approaching from outside and inside of the CNS, respectively. The second order tensor, $n_{ij}$ , accounts for the compliance of the spring layer and this tensor can be expressed in the following form (Qu, 1993; Barai and Weng, 2011):

$$n_{ij} = \alpha\delta_{ij} + (\beta - \alpha)n_i n_j \tag{9}$$

where $n_i$ is the normal outward vector and $\delta_{ij}$ is the Knonecker delta.

A modified expression for the Eshelby's tensor for the case of CNS inclusions with weakened interfaces is written as follows (Qu, 1993; Barai and Weng, 2011):

$$[\bar{S}^E] = [S^E] + ([I] - [S^E])[H][C^m]([I] - [S^E]) \tag{10}$$

in which

$$[H] = [I] + \alpha[P] + (\beta - \alpha)[Q]$$

Elements of  tensors P and Q are given in the Appendix A. Note that the second term in the right hand side of Eq. (10) is produced due to the introduction of the weakened interface. Parameters α and β represent the compliance in the tangential and normal directions respectively as shown in Fig. 4.

Once the modified Eshelby's tensor has been obtained, the expression for the modified Mori-Tanaka estimate for the nano-engineered interphase is derived as follows:

$$[C] = \left(v_m[C^m] + v_{CNS}[C^{CNS}][\tilde{A}_3]\right)\left(v_m[I] + v_{CNS}\left[[\tilde{A}_3] + [H][C^{CNS}][\tilde{A}_3]\right]\right)^{-1} \tag{11}$$

in which

$$[\tilde{A}_3] = \left[[I] + [\bar{S}^E]([C^m])^{-1}([C^{CNS}] - [C^m])\right]^{-1}$$

## 3. Three-phase pull-out model

The axisymmetric RVE of the multiscale composite consists of a microscale fiber embedded in a compliant matrix having a nano-engineered interphase between them as illustrated in Fig. 3. Each constituent of the composite is regarded as a transversely isotropic linear elastic continuum. The cylindrical coordinate system (r–θ–z) is considered in such a way that the axis of the representative volume element (RVE) coincides with the z–axis. The RVE of a multiscale composite has the radius c and the length $L_f$; the radius and the length of the microscale fiber are denoted by a and $L_f$, respectively; the inner and outer radii of the interphase are a and b, respectively. Within the nano-engineered interphase, CNS are randomly or orderly distributed in the epoxy matrix. Both interphase and epoxy matrix are fixed at one end (z = 0) and a tensile stress, $\sigma_p$, is applied on the other end (z = $L_f$) of the embedded microscale fiber. Analytical solutions are obtained for free boundary conditions at the external surface of the matrix cylinder to model a single fiber pull-out problem. The mode of deformation is axisymmetric and thus the stress components ($\sigma_r$, $\sigma_\theta$, $\sigma_z$ and $\sigma_{rz}$) and the displacement components (w, u) are in all three phases function of r and z only. For an axisymmetric geometry, with cylindrical coordinates r, θ, and z, the governing equilibrium equations in terms of normal and the shear stresses are given by:

$$\frac{\partial\sigma_r^{(k)}}{\partial r} + \frac{\partial\sigma_{rz}^{(k)}}{\partial z} + \frac{\sigma_r^{(k)} - \sigma_\theta^{(k)}}{r} = 0 \tag{12}$$

$$\frac{\partial\sigma_z^{(k)}}{\partial z} + \frac{1}{r}\frac{\partial(r\sigma_{rz}^{(k)})}{\partial r} = 0 \tag{13}$$

while the relevant constitutive relations are

$$\begin{Bmatrix} \sigma_r^{(k)} \\ \sigma_\theta^{(k)} \\ \sigma_z^{(k)} \\ \sigma_{rz}^{(k)} \end{Bmatrix} = \begin{bmatrix} C_{11}^{(k)} & C_{12}^{(k)} & C_{13}^{(k)} & 0 \\ C_{12}^{(k)} & C_{22}^{(k)} & C_{23}^{(k)} & 0 \\ C_{13}^{(k)} & C_{23}^{(k)} & C_{33}^{(k)} & 0 \\ 0 & 0 & 0 & C_{66}^{(k)} \end{bmatrix} \begin{Bmatrix} \epsilon_r^{(k)} \\ \epsilon_\theta^{(k)} \\ \epsilon_z^{(k)} \\ \epsilon_{rz}^{(k)} \end{Bmatrix} \quad ; k = f, i \text{ and } m \tag{14}$$

in which the superscripts f, i and m denote the microscale fiber, the interphase and the matrix, respectively. For the $k^{th}$ constituent phase, $\sigma_z^{(k)}$ and $\sigma_r^{(k)}$ represent the normal stresses in the z and r directions, respectively; $\epsilon_z^{(k)}$, $\epsilon_\theta^{(k)}$ and $\epsilon_r^{(k)}$ are the normal strains along the z, θ and r, directions, respectively; $\sigma_{rz}^{(k)}$ is the transverse shear stress, $\epsilon_{rz}^{(k)}$ is the transverse shear strain and $C_{ij}^{(k)}$ are the elastic constants. The strain-displacement relations for an axisymmetric problem relevant to this RVE are

$$\epsilon_z^{(k)} = \frac{\partial w^{(k)}}{\partial z}, \epsilon_\theta^{(k)} = \frac{u^{(k)}}{r}, \epsilon_r^{(k)} = \frac{\partial u^{(k)}}{\partial r} \text{ and } \epsilon_{rz}^{(k)} = \frac{\partial w^{(k)}}{\partial r} + \frac{\partial u^{(k)}}{\partial z} \tag{15}$$

in which $w^{(k)}$and $u^{(k)}$ represent the axial and radial displacements at any point of the $k^{th}$ phase along the z and r directions, respectively. The interfacial traction continuity conditions are given by

$$\sigma_r^f\big|_{r=a} = \sigma_r^i\big|_{r=a};\quad \sigma_{rz}^f\big|_{r=a} = \sigma_{rz}^i\big|_{r=a} = \tau_1;\ \sigma_r^i\big|_{r=b,} = \sigma_r^m|_{r=b};\ \sigma_{rz}^i\big|_{r=b} = \sigma_{rz}^m|_{r=b} = \tau_2;$$

$$w^f\big|_{r=a} = w^i\big|_{r=a};\quad w^i\big|_{r=b} = w^m|_{r=b};\ u^f\big|_{r=a} = u^i\big|_{r=a} \text{ and } u^i\big|_{r=b} = u^m|_{r=b} \tag{16}$$

where $\tau_1$ is the transverse shear stress at the interface between the microscale fiber and the interphase while $\tau_2$ is the transverse shear stress at the interface between the interphase and the matrix. The average axial stresses in the different phases are defined as

$$\bar{\sigma}_z^f = \frac{1}{\pi a^2}\int_0^a \sigma_z^f 2\pi r\, dr;$$

$$\bar{\sigma}_z^i = \frac{1}{\pi(b^2 - a^2)}\int_a^b \sigma_z^i 2\pi r\, dr \quad \text{and } \bar{\sigma}_z^m = \frac{1}{\pi(c^2 - b^2)}\int_b^c \sigma_z^m 2\pi r\, dr \tag{17}$$

Now, making use of Eqs. (12), (13), (16) and (17), it can be shown that

$$\frac{\partial \bar{\sigma}_z^f}{\partial z} = -\frac{2}{a}\tau_1;\ \frac{\partial \bar{\sigma}_z^i}{\partial z} = \frac{2a}{b^2 - a^2}\tau_1 - \frac{2b}{b^2 - a^2}\tau_2 \text{ and } \frac{\partial \bar{\sigma}_z^m}{\partial z} = \frac{2b}{c^2 - b^2}\tau_2 \tag{18}$$

Using the equilibrium equation given by Eq. (13), the transverse shear stresses in the interphase and the matrix can be expressed in terms of the interfacial shear stresses $\tau_1$and $\tau_2$, respectively, as follows:

$$\sigma_{rz}^i = \frac{a}{r}\tau_1 + \frac{1}{2r}(a^2 - r^2)\left\{\frac{2a}{b^2 - a^2}\tau_1 - \frac{2b}{b^2 - a^2}\tau_2\right\} \tag{19}$$

$$\sigma_{rz}^m = \left(\frac{c^2}{r} - r\right)\frac{b}{c^2 - b^2}\tau_2 + \frac{c}{r}\tau \tag{20}$$

Also, since the RVE is an axisymmetric problem, it is usually assumed (Nairn, 1997) that the gradient of the radial displacements with respect to the z–direction is negligible and so, from the constitutive relation given by Eq. (14) and the strain-displacement relations given by Eq. (15) between $\sigma_{rz}^{(k)}$ and $\epsilon_{rz}^{(k)}$, we can write

$$\frac{\partial w^{(k)}}{\partial r} \approx \frac{1}{C_{66}^{(k)}} \sigma_{rz}^{(k)} \; ; \text{ i and m} \tag{21}$$

Solving Eq. (21) and enforcing the continuity conditions at r = a and r = b, respectively, the axial displacements of the interphase and the matrix along the z–direction can be derived as follows:

$$w^i = w_a^f + A_1\tau_1 + A_2\tau_2 \tag{22}$$

$$w^m = w_a^f + A_3\tau_1 + A_4\tau_2 + \frac{c}{C_{66}^m} \tau \ln\left(\frac{r}{b}\right) \text{ and} \tag{23}$$

$$w_a^f = w^f\big|_{r=a} \tag{24}$$

in which $A_i$ (i = 1, 2, 3 and 4) are the constants of the displacement fields and are explicitly given in Appendix B. All other constants, $A_i$, obtained in the course of deriving the pull-out model herein are also explicitly expressed in Appendix B.

The radial displacements in the three constituent phases can be assumed as (Hashin, 1964)

$$u^f = C_1 r, \quad u^i = A_i r + \frac{B_i}{r} \text{ and } u^m = C_2 r + \frac{C_3}{r} \tag{25}$$

where $C_1$, $A_i$, $B_i$, $C_2$ and $C_3$ are the unknown constants. Invoking the continuity conditions for the radial displacement at the interfaces r = a and b, the radial displacement in the interphase can be augmented as follows:

$$u^i = \frac{a^2}{b^2 - a^2}\left(\frac{b^2}{r} - r\right) C_1 - \frac{b^2}{b^2 - a^2}\left(\frac{a^2}{r} - r\right) C_2 - \frac{1}{b^2 - a^2}\left(\frac{a^2}{r} - r\right) C_3 \tag{26}$$

Using displacement fields, constitutive relations and strain-displacement relations, the expressions for the normal stresses can be written in terms of the unknown constants $C_1$, $C_2$ and $C_3$ as follows:

$$\overline{\sigma}_z^f = C_{11}^f \frac{\partial w_a^f}{\partial z} + 2C_{12}^f C_1 \tag{27}$$

$$\sigma_r^f = \frac{C_{12}^f}{C_{11}^f}\bar{\sigma}_z^f + \left[C_{23}^f + C_{33}^f - \frac{2\left(C_{12}^f\right)^2}{C_{11}^f}\right]C_1 \tag{28}$$

$$\sigma_z^i = \frac{C_{11}^i}{C_{11}^f}\bar{\sigma}_z^f - \left(\frac{2C_{12}^i a^2}{b^2 - a^2} + \frac{2C_{12}^f C_{11}^i}{C_{11}^f}\right)C_1 + \frac{2C_{12}^i b^2}{b^2 - a^2}C_2 + \frac{2C_{12}^i}{b^2 - a^2}C_3 + C_{11}^i A_1 \frac{\partial \tau_1}{\partial z}$$

$$+C_{11}^i A_2 \frac{\partial \tau_2}{\partial z} \tag{29}$$

$$\sigma_r^i = \frac{C_{13}^i}{C_{11}^f}\bar{\sigma}_z^f + \left[\frac{C_{23}^i a^2}{b^2 - a^2}\left(\frac{b^2}{r^2} - 1\right) + \frac{C_{33}^i a^2}{b^2 - a^2}\left(-\frac{b^2}{r^2} - 1\right) - \frac{2C_{12}^f C_{13}^i}{C_{11}^f}\right]C_1$$

$$+\left[-\frac{C_{23}^i b^2}{b^2 - a^2}\left(\frac{a^2}{r^2} - 1\right) + \frac{C_{33}^i b^2}{b^2 - a^2}\left(\frac{a^2}{r^2} + 1\right)\right]C_2$$

$$+\left[-\frac{C_{23}^i}{b^2 - a^2}\left(\frac{a^2}{r^2} - 1\right) + \frac{C_{33}^i}{b^2 - a^2}\left(\frac{a^2}{r^2} + 1\right)\right]C_3 + C_{13}^i A_1 \frac{\partial \tau_1}{\partial z} + C_{13}^i A_2 \frac{\partial \tau_2}{\partial z} \tag{30}$$

$$\sigma_z^m = \frac{C_{11}^m}{C_{11}^f}\bar{\sigma}_z^f - \frac{2C_{12}^f C_{11}^m}{C_{11}^f}C_1 + 2C_{12}^m C_2 + C_{11}^m A_3 \frac{\partial \tau_1}{\partial z} + C_{11}^m A_4 \frac{\partial \tau_2}{\partial z} \tag{31}$$

$$\sigma_r^m = \frac{C_{12}^m}{C_{11}^f}\bar{\sigma}_z^f - \frac{2C_{12}^f C_{12}^m}{C_{11}^f}C_1 + (C_{11}^m + C_{12}^m)C_2 + (C_{12}^m - C_{11}^m)\frac{C_3}{r^2} + C_{12}^m A_3 \frac{\partial \tau_1}{\partial z} + +C_{12}^m A_4 \frac{\partial \tau_2}{\partial z} \tag{32}$$

Invoking the continuity conditions $\sigma_r^f\big|_{r=a} = \sigma_r^i\big|_{r=a}$ and $\sigma_r^i\big|_{r=b} = \sigma_r^m\big|_{r=b}$, the following equations for solving unknown constants $C_1$, $C_2$ and $C_3$ are obtained:

$$\begin{bmatrix} B_{11} & B_{12} & B_{13} \\ B_{21} & B_{22} & B_{23} \\ B_{31} & B_{32} & B_{33} \end{bmatrix}\begin{Bmatrix} C_1 \\ C_2 \\ C_3 \end{Bmatrix} = \frac{\bar{\sigma}_z^f}{C_{11}^f}\begin{Bmatrix} C_{12}^f - C_{13}^i \\ C_{13}^i - C_{12}^m \\ -C_{12}^m \end{Bmatrix} + \begin{Bmatrix} 0 \\ A_5 - A_7 \\ -C_{12}^m A_3 \end{Bmatrix}\frac{\partial \tau_1}{\partial z} + \begin{Bmatrix} 0 \\ A_6 - A_8 \\ -A_9 \end{Bmatrix}\frac{\partial \tau_2}{\partial z} \tag{33}$$

The expressions of the coefficients $B_{ij}$ are presented in Appendix B. Solving Eq. (33), the solutions of the constants $C_1$, $C_2$ and $C_3$ can be expressed as:

$$C_i = b_{i1}\bar{\sigma}_z^f + b_{i2}\frac{\partial \tau_1}{\partial z} + b_{i3}\frac{\partial \tau_2}{\partial z}; \quad i = 1, 2, 3 \tag{34}$$

The expressions of the coefficients $b_{i1}$, $b_{i2}$ and $b_{i3}$ are evident from Eq. (34), and are not shown here for the sake of clarity. Now, making use of Eqs. (29), (31) and (34) in the last two equations of (20), respectively, the average axial stresses in the interphase and the matrix are written as follows:

$$\overline{\sigma}_z^i = A_{14}\overline{\sigma}_z^f + A_{15}\frac{\partial\tau_1}{\partial z} + A_{16}\frac{\partial\tau_2}{\partial z} \tag{35}$$

$$\overline{\sigma}_z^m = A_{17}\overline{\sigma}_z^f + A_{18}\frac{\partial\tau_1}{\partial z} + A_{19}\frac{\partial\tau_2}{\partial z} \tag{36}$$

Now, satisfying the equilibrium of force along the axial (z) direction at any transverse cross-section of the RVE, the following equation is obtained:

$$\pi a^2\sigma_p = \pi a^2\overline{\sigma}_z^f + \pi(b^2 - a^2)\overline{\sigma}_z^i + \pi(c^2 - b^2)\overline{\sigma}_z^m \tag{37}$$

where $\sigma_p$ is the pull-out stress applied on the fiber end.

Differentiating the first and last equations of (18) with respect to z, we have

$$\tau_1' = -\frac{a}{2}\frac{\partial^2\overline{\sigma}_z^f}{\partial z^2} \text{ and } \tau_2' = \frac{c^2 - b^2}{2b}\frac{\partial^2\overline{\sigma}_z^m}{\partial z^2} \tag{38}$$

Using Eqs. (35–38), the governing equation for the average axial stress in the microscale fiber is obtained as follows:

$$\frac{\partial^4\overline{\sigma}_z^f}{\partial z^4} + A_{24}\frac{\partial^2\overline{\sigma}_z^f}{\partial z^2} + A_{25}\overline{\sigma}_z^f - A_{26}\sigma_p = 0 \tag{39}$$

Solution of Eq. (39) is given by:

$$\overline{\sigma}_z^f = A_{27}\sinh(\alpha z) + A_{28}\cosh(\alpha z) + A_{29}\sinh(\beta z) + A_{30}\cosh(\beta z) + (A_{26}/A_{25})\sigma_p \tag{40}$$

where

$$\alpha = \sqrt{1/2\left(-A_{24} + \sqrt{(A_{24})^2 - 4A_{25}}\right)} \text{ and } \beta = \sqrt{1/2\left(-A_{24} - \sqrt{(A_{24})^2 - 4A_{25}}\right)} \tag{41}$$

Substitution of Eq. (40) into the first equation of (18) yields the expression for the microscale fiber/interphase interfacial shear stress as follows:

$$\tau_1 = -\frac{a}{2}[A_{27}\alpha\cosh(\alpha z) + A_{28}\alpha\sinh(\alpha z) + A_{29}\beta\cosh(\beta z) + A_{30}\beta\sinh(\beta z)] \tag{42}$$

The stress boundary conditions of the model are given by

$$\bar{\sigma}_z^f\big|_{z=0} = 0, \quad \bar{\sigma}_z^f\big|_{z=L_f} = \sigma_p, \quad \left.\frac{\partial \bar{\sigma}_z^f}{\partial z}\right|_{z=0} = 0 \quad \text{and} \quad \left.\frac{\partial \bar{\sigma}_z^f}{\partial z}\right|_{z=L_f} = 0 \tag{43}$$

$$\bar{\sigma}_z^m|_{z=0} = \frac{a^2\sigma_p}{c^2 - b^2}, \quad \bar{\sigma}_z^m|_{z=L_f} = 0, \quad \left.\frac{\partial \bar{\sigma}_z^m}{\partial z}\right|_{z=0} = 0 \quad \text{and} \quad \left.\frac{\partial \bar{\sigma}_z^m}{\partial z}\right|_{z=L_f} = 0 \tag{44}$$

Utilizing the boundary conditions given by Eq. (43) in Eq. (40), the constants $A_{27}$, $A_{28}$, $A_{29}$ and $A_{30}$ can be obtained. Similar governing equations for the average axial stress in the matrix and the interphase/matrix interfacial shear stress can be derived using the same solution methodology; such that,

$$\frac{\partial^4 \bar{\sigma}_z^m}{\partial z^4} + A_{36}\frac{\partial^2 \bar{\sigma}_z^m}{\partial z^2} + A_{37}\bar{\sigma}_z^m - A_{38}\sigma_p = 0 \tag{45}$$

Solution of Eq. (45) is given by:

$$\bar{\sigma}_z^m = A_{39}\sinh(\alpha_m z) + A_{40}\cosh(\alpha_m z) + A_{41}\sinh(\beta_m z) + A_{42}\cosh(\beta_m z) + (A_{38}/A_{37})\sigma_p \tag{46}$$

where

$$\alpha_m = \sqrt{1/2\left(-A_{36} + \sqrt{(A_{36})^2 - 4A_{37}}\right)} \text{ and } \beta_m = \sqrt{1/2\left(-A_{36} - \sqrt{(A_{36})^2 - 4A_{37}}\right)} \tag{47}$$

Substitution of Eq. (46) into the last equation of (18) yields the expression for the interphase/bulk matrix interfacial shear stress as follows:

$$\tau_2 = \frac{c^2 - b^2}{2b}\left[A_{39}\alpha_m\cosh(\alpha_m z) + A_{40}\alpha_m\sinh(\alpha_m z) + A_{41}\beta_m\cosh(\beta_m z) + A_{42}\beta_m\sinh(\beta_m z)\right] \tag{48}$$

Utilizing the boundary conditions given by Eq. (34) in Eq. (46), the constants $A_{39}$, $A_{40}$, $A_{41}$ and $A_{42}$ can be obtained.

## 4. Results and discussion

In this section, the results of the developed pull-out model are compared with the FE results. Subsequently, the effect of orientation of A-CNT bundles and their loading on the stress transfer behavior of the multiscale composite are analysed and discussed. Considering the epoxy as the matrix phase and the CNS as the reinforcement, the effective elastic properties of the nanocomposite have been determined by following the micromechanical modeling approach developed in section 2.2. Note that the effective elastic properties of nano-engineered interphase are obtained by MD simulations in conjunction with the MT homogenization scheme and are summarized in Table 4. It should be noted that the CNT volume fraction used in the computations (see Table 4) is either with respect to the CNS volume or interphase volume. Unless otherwise stated: (i) the geometrical parameters a, b and c of the pull-out model considered here are as 5 μm, 6.5 μm and 9 μm, respectively; (ii) the value of $L_f$ for the model as 100 μm; and (iii) the CNS, with 5% loading in the interphase, are considered to be aligned with the axis of a microscale fiber and perfectly bonded with the surrounding epoxy. Also the following transversely isotropic elastic properties of the carbon fiber (Honjo, 2007) are used for both analytical and FE analyses presented herein.

$C_{11}$= 236.4 GPa, $C_{12}$= 10.6 GPa, $C_{13}$= $C_{12}$, $C_{22}$= 24.8 GPa, $C_{33}$= $C_{22}$, $C_{44}$= 7 GPa, $C_{55}$= 25 GPa, and $C_{66}$= $C_{55}$.

## *4.1. Validation of the analytical model by FE method*

The novelty of the three-phase pull-out model derived in this study is that the radial as well as the axial deformations of the different transversely isotropic constituent phases of the multiscale composite have been taken into account. Therefore, it is desirable to justify the validity of the pull-out model derived in section 3. For this purpose, 3D axisymmetric FE model has been developed using the commercial software ANSYS 14.0. The geometry of the FE model, and the loading and boundary conditions are chosen such that they represent those of the actual experimental test. Also the geometric and material properties used in the FE simulations are identical to those of the analytical model. The microscale fiber, the interphase and the epoxy matrix are constructed and meshed with twenty-node solid elements SOLID186. Identical to the analytical model, a uniformly distributed tensile stress ($\sigma_p$) is applied to the free end of the microscale fiber along the z–direction at $z = L_f$. The boundary conditions are imposed in such way that the bottom cylindrical surface of the matrix is fixed ($z = 0$) and the top cylindrical surface of the matrix is traction-free. Moreover, the lateral cylindrical surface of the matrix is assumed to be traction-free, to simulate single fiber pull-out problem. The case of fixed lateral surface of matrix to approximately model a hexagonal array of fibers in the matrix is not considered here as the results are identical to that of the single fiber pull-out model as demonstrated by Upadhyaya and Kumar (2015). Three- and two-phase FE models have been developed to validate the analytical pull-out model. The three-phase model made of the microscale fiber, the interphase containing CNS aligned along the direction of a microscale fiber (see Fig. 3), and the matrix; the two-phase model consists of only the microscale fiber and matrix. At first, a mesh convergence study is performed to see the effect of element size on the stress transfer characteristics. Once the convergence is assured, FE simulations are carried out to validate the analytical model. Fig. 5 shows the comparisons of the average axial stress in the microscale fiber computed by the analytical pull-out model and the FE pull-out model. It may be observed that the axial stress decreases monotonously from the pulled fiber end to the embedded fiber end. Moreover, the results predicted by the analytical models are in great accordance with those of FE simulations. As is well known, within the framework of linear theory of elasticity, singular stress field may arise at the corners of bimaterial interfaces between different phases with different elastic properties (Marotzke, 1994). FE results seem to indicate the existence of a singularity at the free surface between the fiber and interphase at the fiber entry. Therefore, in order to ensure realistic stress conditions, i.e. an unrestricted variation of the stress field near the fiber entry, we determined and compared the transverse shear stresses ($\sigma_{rz}^{i}$) in the interphase over a distance of 0.02 fiber diameter away from the fiber-interphase interface. This technique has been adopted by several researchers to ascertain the accuracy of the analytical model (Marotzke, 1994; Upadhyaya and Kumar (2015) and references therein). It may be observed from Fig. 6 that the analytical model provides a good estimation of the transverse shear stress distribution. It may be importantly observed from Fig. 6 that the incorporation of interphase significantly reduces the maximum value of $\sigma_{rz}^{i}$ compared to that of the two-phase model of the composite. Our comparisons clearly indicate that the derived pull-out model herein captures the crucial stress transfer mechanisms of the multiscale composite; therefore, the subsequent results presented herein are based on the analytical pull-out model.

### *4.2. Theoretical results*

In this section, results of parametric study have been presented and the effects of orientations of CNS in the interphase on the stress transfer characteristics of the multiscale composite are discussed. The maximum CNT volume fraction considered in the interphase is 5%, so the CNT volume fraction in the composite is much less than 5%. Three different cases have been considered: (i) interphase containing aligned CNS (parallel), (ii) interphase containing

CNS, which are radially aligned to the axis of a microscale fiber (perpendicular), and (iii) interphase containing randomly dispersed CNS (random). These cases represent the practical situation of distribution of CNS in the epoxy matrix and may significantly influence the overall properties of both interphase and multiscale composite. Fig. 7 demonstrates that the orientations of CNS do not much influence the average axial stress in the microscale fiber. On the other hand, the effect of orientations of CNS is found to be significantly influence the interfacial shear stress (see Fig. 8, which shows $\tau_1$ at the interface between the microscale fiber and the interphase). It may be observed that the maximum value of $\tau_1$ is reduced by 15.5% for the parallel case compared to homogenous interphase. This is attributed to the fact that the axial elastic properties of the interphase are higher in the parallel case (see Table 4) and hence the interfacial properties of the resulting multiscale composite improve in comparison to all other cases. It may also be observed from Fig. 8 that the perpendicular case does not improve the interfacial properties of the multiscale composite; but the random case marginally enhances the interfacial properties of the multiscale composite over the pure matrix case. This is due to the fact that the axial elastic properties of the interphase in the perpendicular case are matrix dominant and hence its results are found to be more close to those of the pure matrix case; in the random case, CNS are homogeneously dispersed in the interphase and hence its overall elastic properties improve in comparison to the pure matrix case.

Fig. 9 demonstrates the radial stress at the fiber-interphase interface along the length of the fiber for the different cases of orientation of CNS. Here, $r_i = 5$ μm and 6.5 μm represent the fiber-interphase interface and the interphase-matrix interface, respectively. It is evident from this figure that the radial stress peaks at the loaded fiber end and decays rapidly with increasing distance from it, changing from the tensile to compressive. This figure also demonstrates that the magnitude is nearly equal at both the entry and exit ends of the fiber because of the small aspect ratio of the microscale fiber. Radial stress at the interface at the fibre entry is tensile due to Poisson’s contraction of the fibre, while at the exit end it is compressive due to Poisson’s contraction of the matrix; this is consistent with the trend observed in several studies (Quek and Yue, 1997; Upadhyaya and Kumar (2015) and references therein). Such a concentrated stress distribution close to stress singular point makes the interface region very susceptible to mixed mode fracture. Among all the cases shown in Fig. 9, the parallel case reduces the maximum value of radial stress at the loaded fiber end by 42% compared to the pure matrix case. This is very important finding, because the debonding failure for small fiber aspect ratios is mode I dominant, and such debonding failure can be prevented by nano-engineered interphases. Fig. 10 shows the transverse shear stress distribution in the rz plane of the interlayer. Closely observing the slope of shear stress curve in different regions, it is seen that the value of shear stress increases rapidly towards the other interface at $z = 0$; this is true for all the four cases but the parallel case provides better results. It may be observed from Fig. 10 that the magnitude of radial stress is relatively lower compared to the magnitude of shear stress. Radial tension at fiber-matrix interface may initiate debonding at the loaded fiber end if the radial stress exceeds the interfacial tensile strength and such microscale damages can grow rapidly leading to macroscale failure of the component (Upadhyaya and Kumar, 2015). The large reduction in the radial stress at the interface indicates that the nano-engineered interphase significantly improves the mode I fracture toughness of a resulting composite, which is likely to affect the initial mode of failure of the composite.

So far, in this work, the stress transfer characteristics of the multiscale composite have been studied by considering the four different cases. It is evident from the previous set of results that the parallel case improves the stress transfer behavior of multiscale composite; therefore, we consider the parallel case to further study the role of loading of CNTs on the stress transfer characteristics of the multiscale composite. Practically, the CNT loading in the interphase can vary around the microscale fiber, therefore, the investigation of the effect of variation of CNT loading on the stress transfer characteristics of the multiscale composite would be an important study. For such investigation, three discrete values of CNT loading are considered: 5%, 10% and 15%. As can be seen from the tabulated values in Table 4, the elastic properties of the interphase improve with the increase in the CNT loading. Once again it is found that the average axial stress in the microscale fiber is not much influenced by the loading of CNTs in comparison to the base composite, i.e., without CNTs, as demonstrated in Fig. 11. Fig. 12 demonstrates that the effect of CNT loading on the interfacial shear stress between the microscale fiber and the interphase. This figure indicates that the incorporation of different types of interphase exhibit almost identical trends and the loading of CNTs more than 5% do not much reduce the maximum value of $\tau_1$. On the other hand, the effect of CNT loading on the radial stress at the fiber-interphase interface along the length of the microscale fiber is found to be marginal, as shown in Fig. 13. Note that the transverse elastic coefficients of interphase dominate the constants$A_{27}$, $A_{28}$, $A_{29}$ and $A_{30}$ appearing in Eq. (42). We can observe from Table 4 that the transverse elastic coefficients of interphase containing aligned CNS are not significantly improved at higher CNT volume fractions. Therefore, CNT volume fraction does not much influence the interfacial shear stress when CNS are parallel to the microscale fiber.

Next, we consider the effect of imperfect CNS-epoxy interfacial boding on the stress transfer characteristics of the multiscale composite, and once again consider the parallel case with 5% CNT loading. The effective elastic properties of the interphase have been determined by following the micromechanical modeling approach developed in section 2.2.1 and are summarized in Table 5. Two values of sliding parameter (α) are considered, $0.5 \times 10^{-6}\,\mathrm{nm/GPa}$ and $0.75 \times 10^{-6}\,\mathrm{nm/GPa}$, and $\beta = 0$. When $\beta = 0$, the CNS-epoxy interface is allowed to slide without normal separation or interpenetration. We can observe from the tabulated values in Table 5 that the transverse elastic properties of the interphase degrade as the value of α increase. This finding is consistent with the previously reported results (Esteva and Spanos, 2009; Barai and Weng, 2011; Pan et al., 2013). Figure 14 demonstrates that the effect of imperfect interfacial bonding on the average axial stress in the microscale fiber is not significant. Although not presented here, the same is true for the interfacial shear stress along the length of the microscale fiber. On the other hand, Fig. 15 reveals that the imperfect interfacial bonding significantly affects the radial stress at the fiber-interphase interface along the length of the microscale fiber to such extent that the influence of CNS diminishes completely and renders the engineered interphase to behave in the same manner as that of pure matrix.

## 5. Conclusions

In this study, we proposed a novel concept to improve the mechanical properties and stress transfer behaviour of a multiscale composite. This concept involves the introduction of carbon nanostructures (CNS) around the microscale fibers embedded in the epoxy matrix, resulting in a multiscale composite with enhanced properties. Using this concept, damage initiation threshold and the fatigue strength of conventional composites can be greatly improved by toughening the interfacial fiber-matrix region. Accordingly, two aspects are examined: (i) determination of the transversely isotropic properties of CNS composed of A-CNT bundle and epoxy matrix through MD simulations in conjunction with the Mori-Tanaka model, and (ii) the development of three-phase pull-out model for a multiscale composite. Our pull-out model incorporates different nano-and micro-scale transversely isotropic phases and allows the quantitative determination of the stress transfer characteristics of the multiscale composites. This model has also been validated by comparing the predicted results with those of FE simulations. The developed analytical model was then applied to investigate the effect of orientation of CNS considering their perfect and imperfect interfacial bonding conditions with the surrounding epoxy on the stress transfer characteristics of the multiscale composite. The following is a summary of our findings:

1. The incorporation of the interphase – containing CNS and epoxy matrix – between the fiber and the matrix would significantly improve the mode I fracture toughness of a resulting composite, which is likely to affect the initial mode of failure of the composite.
2. Orientation of CNS has significant influence on the stress transfer characteristics of the multiscale composite; aligned CNS along the axial direction of the microscale fiber is found be effective in comparison to all other orientations.
3. CNS-epoxy interface weakening significantly affect the radial stress along the length of the microscale fiber.
4. The three-phase pull-out model developed in this study offers significant advantages over the existing pull-out models and is capable of investigating the stress transfer characteristics of any multiscale composite containing either aligned or randomly dispersed nanostructures.

## Acknowledgement

This work was funded by the Lockheed Martin Corporation (Award no: 13NZZA1). S. K. would like to thank Professor Brian L. Wardle, Massachusetts Institute of Technology, for his helpful suggestion and comments on MS and Dr. Tushar Shah of Lockheed Martin Corporation. The authors also wish to thank the anonymous reviewers for their helpful comments and suggestions.

## References

Ahmed, K.S., Keng, A. K., 2012. A pull-out model for perfectly bonded carbon nanotube in polymer composites. *Journal of Mechanics of Materials and Structures* 7, 753–764. doi: 10.2140/jomms.2012.7.753

Alian, A.R., Kundalwal, S.I., Meguid, S.A., 2015a. Interfacial and mechanical properties of epoxy nanocomposites using different multiscale modeling schemes. *Composite Structures* 131, 545–555. doi: 10.1016/j.compstruct.2015.06.014

Alian, A.R., Kundalwal, S.I., Meguid, S.A., 2015b. Multiscale modeling of carbon nanotube epoxy composites. *Polymer* 70, 149–160 . doi:10.1016/j.polymer.2015.06.004

Allen, M.P., Tildesley, D.J., 1987. *Computer simulation of liquids*. Oxford: Clarendon Press.

Arif, M.F., Kumar, S. and Shah, T., 2016. Tunable morphology and its influence on electrical, thermal and mechanical properties of carbon nanostructure-buckypaper. *Materials & Design*, *101*, pp.236-244. doi:10.1016/j.matdes.2016.03.122

Bandyopadhyay, A., Odegard, G.M., 2012. Molecular modeling of crosslink distribution in epoxy polymers. *Modelling and Simulation in Materials Science Engineering* 20, 045018. doi:10.1088/0965-0393/20/4/045018

Banholzer, B., Brameshuber, W., Jung, W., 2005. Analytical simulation of pull-out tests—the direct problem. *Cement and Concrete Composites* 27, 93–101. doi: 10.1016/j.cemconcomp.2004.01.006

Barai, P., Weng, G.J., 2011. A theory of plasticity for carbon nanotube reinforced composites. *International Journal of Plasticity* 27, 539–559. doi: 10.1016/j.ijplas.2010.08.006

Bekyarova, E., Thostenson, E.T., Yu, A., 2007. Multiscale carbon nanotube-carbon fiber reinforcement for advanced epoxy composites. *Langmuir* 23, 3970–3974. doi:10.1021/la062743p

Benveniste, Y., 1987. A new approach to the application of Mori-Tanaka's theory in composite materials. *Mechanics of Materials* 6(2), 147–157. doi: 10.1016/0167-6636(87)90005-6

Chen, L., Jin, H., Xu, Z., Li, J., Guo, Q., Shan, M., Yang, C., Wang, Z., Mai, W., Cheng, B., 2015. Role of a gradient interface layer in interfacial enhancement of carbon fiber/epoxy hierarchical composites. *Journal of Material Science* 50, 112–121. doi:10.1007/s10853-014-8571-y

Cho, J., Chen, J.Y., Daniel, I.M., 2007. Mechanical enhancement of carbon fiber/epoxy composites by graphite nanoplatelet reinforcement. *Scripta Materialia* 56, 685–688. doi: 10.1016/j.scriptamat.2006.12.038

Cui, Y., Kundalwal, S. I., Kumar S., 2016. Gas barrier performance of graphene/polymer nanocomposites. *Carbon* 98, 313–333. doi:10.1016/j.carbon.2015.11.018

Dauber-Osguthorpe, P., Roberts, V.A., Osguthorpe, D.J., Wolff, J., Genest, M., Hagler, A.T., 1998. Structure and energetics of ligand binding to proteins: Escherichia coli dihydrofolate reductase-trimethoprim, a drug-receptor system. *Proteins* 4, 31–47. doi:10.1002/prot.340040106

Davis, D.C., Wilkerson, J.W., Zhu, J., Ayewah, D.O.O., 2010. Improvements in mechanical properties of a carbon fiber epoxy composite using nanotube science and technology. *Composite Structures* 92, 2653–2662. doi:10.1016/j.compstruct.2010.03.019

Davis, D.C., Wilkerson, J.W., Zhu, J., Hadjiev, V.G., 2011. A strategy for improving mechanical properties of a fiber reinforced epoxy composite using functionalized carbon nanotubes. *Composites Science and Technology* 71, 1089–1097. doi: 10.1016/j.compscitech.2011.03.014

Drzal, L., 1986. *The interphase in epoxy composites*. In: Duek K, editor. Epoxy Resins and Composites II. vol. 75 of Advances in Polymer Science. Springer Berlin Heidelberg, 1–32. doi: 10.1007/BFb0017913

Dumlich, H., Gegg, M., Hennrich, F., Reich, S., 2011. Bundle and chirality influences on properties of carbon nanotubes studied with van der Waals density functional theory. *Physica Status Solidi B* 248, 2589–2592. doi: 10.1002/pssb.201100212

Esteva, M., Spanos, P. D., 2009. Effective elastic properties of nanotube reinforcedcomposites with slightly weakened interfaces. *Journal of Mechanics of Materials and Structures* 4(5), 887–900. doi: 10.2140/jomms.2009.4.887

Frankland, S.J.V., Harik, V.M., Odegard, G.M., Brenner, D.W., Gates, T.S., 2003. The stress-strain behavior of polymer-nanotube composites from molecular dynamics simulation. *Composites Science and Technology* 63(11), 1655–61. 10.1016/S0266-3538(03)00059-9

Fu, S.-Y., Lauke, B., 2000. Comparison of the stress transfer in single- and multi-fiber composite pull-out tests. *Journal of Adhesion Science and Technology* 14(3), 437–452. doi: 10.1163/156856100742690

Fu, S.-Y., Yue, C.-Y., Hu, X., Mai, Y.-W., 2000. Analyses of the micromechanics of stress transfer in single- and multi-fiber pull-out tests. *Composites Science and Technology* 60, 569–579. doi: 10.1016/S0266-3538(99)00157-8

Garcia, E.J., Wardle, B.L., Hart, A.J., Yamamoto, N., 2008. Fabrication and multifunctional properties of a hybrid laminate with aligned carbon nanotubes grown In Situ. *Composites Science and Technology* 68, 2034–2041. doi:10.1016/j.compscitech.2008.02.028

Haghighatpanah, S., Bolton, K., 2013. Molecular-level computational studies of single wall carbon nanotube–polyethylene composites. *Computational Materials Science* 69, 443–454. doi: 10.1016/j.commatsci.2012.12.012

Hashin, Z., Rosen, B. W., 1964. The elastic moduli of fiber-reinforced materials. *ASME Journal Applied Mechanics* 31(2), 223–232. doi: 10.1115/1.3629590

Honjo, K., 2007. Thermal stresses and effective properties calculated for fiber composites using actual cylindrically-anisotropic properties of interfacial carbon coating. *Carbon* 45(4), 865–872. 10.1016/j.carbon.2006.11.007

Hung, K.H., Kuo, W.S., Ko, T.H., Tzeng, S.S., Yan, C.F., 2009. Processing and tensile characterization of composites composed of carbon nanotube-grown carbon fibers. *Composites Part A* 40, 1299–1304. doi:10.1016/j.compositesa.2009.06.002

Jia, Y., Chen, Z., Yan, W., 2014. A numerical study on carbon nanotube–hybridized carbon fibre pullout. *Composites Science and Technology* 91, 38–44. doi: 10.1016/j.compscitech.2013.11.020

Jiang, Q., Tallury, S.S., Qiu, Y., Pasquinelli, M.A., 2014. Molecular dynamics simulations of the effect of the volume fraction on unidirectional polyimide–carbon nanotube nanocomposites. *Carbon* 67, 440–448. doi: 10.1016/j.carbon.2013.10.016

Jin, S.-Y., Young, R.J., Eichhorn, S.J. Hybrid carbon fibre–carbon nanotube composite interfaces *Composites Science and Technology* 95, 114–120. doi: 10.1016/j.compscitech.2014.02.015

Kim, J.K., Baillie, C., Mai, Y.-W., 1992. Interfacial debonding and fiber pull-out stresses. Part I. Critical comparison of existing theories with experiments. *Journal of Material Science* 27, 3143–3154. doi:10.1007/BF01116004

Kim, J.K., Lu, S., Mai, Y.-W., 1994. Interfacial debonding and fiber pullout stresses. Part IV. Influence of interface layer on the stress transfer. *Journal of Material Science* 29, 554–561. doi: 10.1007/BF01162521

Kumar, A., Sundararaghavan, V., Browning, A.R., 2014. Study of temperature dependence of thermal conductivity in cross-linked epoxies using molecular dynamics simulations with long range interactions. *Modelling and Simulation in Materials Science and Engineering* 22, 025013. doi: 10.1088/0965-0393/22/2/025013

Kumar, V.R., Kumar, S., Pal, G. and Shah, T., 2016. High-Ampacity Overhead Power Lines With Carbon Nanostructure–Epoxy Composites. *Journal of Engineering Materials and Technology*, *138*(4), p.041018. doi: 10.1115/1.4034095

Kundalwal, S.I., Meguid, S.A., 2015. Micromechanics modelling of the effective thermoelastic response of nano-tailored composites. *European Journal of Mechanics – A/Solids* 53, 241–253. doi: 10.1016/j.euromechsol.2015.05.008

Kundalwal, S.I., Ray, M.C., Meguid, S.A., 2014a. Shear lag model for regularly staggered short fuzzy fiber reinforced composite. *ASME Journal of Applied Mechanics* 81(9), 091001-1–091001-14. doi: 10.1115/1.4027801

Kundalwal, S.I., Ray, M.C., 2014b. Shear lag analysis of a novel short fuzzy fiber-reinforced composite. *Acta Mechanica* 225(9), 2621-2643. doi: 10.1007/s00707-014-1095-3

Kundalwal, S.I., Kumar, R.S., Ray, M.C., 2014c. Effective thermal conductivities of a novel fuzzy carbon fiber heat exchanger containing wavy carbon nanotubes. *International Journal of Heat and Mass Transfer* 72, 440-451. doi: 10.1016/j.ijheatmasstransfer.2014.01.025

Lachman, N., Wiesel, E., de Villoria R.G., Wardle, B.L., Wagner, H.D., 2012. Interfacial load transfer in carbon nanotube/ceramic microfiber hybrid polymer composites. *Composites Science and Technology* 72, 1416–1422. doi:10.1016/j.compscitech.2012.05.015

Li, C., Medvedev, G.A., Lee, E.-W., Kim, J., Caruthers, J.M., Strachan, A., 2012. Molecular dynamics simulations and experimental studies of the thermomechanical response of an epoxy thermoset. *Polymer* 53, 4222–4230. doi:10.1016/j.polymer.2012.07.026

Li, C., Strachan, A., 2010. Molecular simulations of crosslinking process of thermosetting polymers. *Polymer* 51**,** 6058–6070. doi: 10.1016/j.polymer.2010.10.033

Liljenhjerte, J., Kumar, S., 2015. Pull-Out Performance of 3D Printed Composites with Embedded Fins on the Fiber. *MRS Proceedings*. Vol. 1800. Cambridge University Press.

Marotzke, C., 1994. The elastic field arising in the single-fiber pull-out test. *Composites Science and Technology* 50(3), 393–405. doi: 10.1016/0266-3538(94)90027-2

Martínez, L., Andrade, R., Birgin, E.G., Martínez, J.M., 2009. PACKMOL: a package for building initial configurations for molecular dynamics simulations. *Journal of Computational Physics* 30**,** 2157–2164. doi: 10.1002/jcc.21224

Marzari, N., Ferrari, M., 1992. Textural and micromorphological effects on the overall elastic response of macroscopically anisotropic composites. *Journal of Applied Mechanics* 59, 269–275. doi: 10.1115/1.2899516

Meng, Q., Wang, Z., 2015. Theoretical model of fiber debonding and pull-out in unidirectional hybrid-fiber-reinforced brittle-matrix composites. *Journal of Composite Materials* 49(14), 1739–1751. doi: 10.1177/0021998314540191

Mori, T., Tanaka, K., 1973. Average stress in matrix and average elastic energy of materials with misfitting inclusions. *Acta Metallurgica* 21(5), 571–574. doi: 10.1016/0001-6160(73)90064-3

Morris, J.E., 2008. In *Nanopackaging: nanotechnologies and electronics packaging*. Springer: Portland, OR, 49–50. doi: 10.1007/978-0-387-47325-3

Nairn, J. A., 1997. On the use of shear-lag methods for analysis of stress transfer in unidirectional composites. *Mechanics of Materials* 26(2), 63–80. doi: 10.1016/S0167-6636(97)00023-9

Odegard, G.M., Gates, T.S., Wise, K.E., Park, C., Siochi, E.J., 2003. Constitutive modeling of nanotube-reinforced polymer composites. *Composites Science and Technology* 2003, 63, 1671–1687. doi:10.1016/S0266-3538(03)00063-0

Pal, G., Kumar, S., 2016a. Multiscale modeling of effective electrical conductivity of short carbon fiber-carbon nanotube-polymer matrix hybrid composites. *Materials and Design*, 89, 129-136. doi:10.1016/j.matdes.2015.09.105

Pal, G., Kumar, S., 2016b. Modeling of carbon-nanotubes and carbon nanotube-polymer composites. *Progress in Aerospace Sciences*. doi:10.1016/j.paerosci.2015.12.001

Pan, Y., Weng, G.J., Meguid, S.A., Bao, W.S., Zhu, Z.H., Hamouda, A.M.S., 2013. Interface effects on the viscoelastic characteristics of carbon nanotube polymer matrix composites. *Mechanics of Materials,* 58, 1-11. doi: 10.1016/j.mechmat.2012.10.015

Pascault, J.P., Sautereau, H., Verdu, J., Williams, R.J.J., 2002. *Thermosetting polymers*. New York: CRC Press. doi: 10.1002/9781118480793.ch28

Plimpton, S., 1995. Fast parallel algorithms for short-range molecular dynamics. *Journal of Computational Physics* 117, 1–19. doi:10.1006/jcph.1995.1039

Qiu, Y.P., Weng, G.J., 1990. On the application of Mori-Tanaka's theory involving transversely isotropic spheroidal inclusions. *International Journal of Engineering Science* 28(11), 1121–1137. doi: 10.1016/0020-7225(90)90112-V

Qu, J., 1993. The effect of slightly weakened interfaces on the overall elastic properties of composite materials. *Mechanics of Materials,* 14(4), 269–281. doi: 10.1016/0167-6636(93)90082-3

Quek, M.Y., Yue, C.Y., 1997. An improved analysis for axisymmetric stress distributions in the single fibre pull-out test. *Journal of Material Science* 37, 5457–5465. doi: 10.1023/A:1018647718400

Ray, M.C., Kundalwal, S.I., 2013a. A thermomechanical shear lag analysis of short fuzzy fiber reinforced composite containing wavy carbon nanotubes. *European Journal of Mechanics – A/Solids* 44, 41-60. doi: 10.1016/j.euromechsol.2013.10.001

Ray, M.C., Kundalwal, S.I., 2013b. Effect of carbon nanotube waviness on the load transfer characteristics of short fuzzy fiber-reinforced composite. ASCE *Journal of Nanomechanics and Micromechanics* 4(2), A4013010. doi: 10.1061/(ASCE)NM.2153-5477.0000082

Romanov, V., Lomov, S. V., Verpoest, I., Gorbatikh, L., 2015. Inter-fiber stresses in composites with carbon nanotube grafted and coated fibers. *Composites Science and Technology* 114, 79–86. doi: 10.1016/j.compscitech.2015.04.013

Sottos, N.R., McCullough, R.L., Scott, W.R., 1992. The influence of interphase regions on local thermal displacements in composites. *Composites Science and Technology* 44(4), 319–332. doi: 10.1016/0266-3538(92)90069-F

Tsai, J.L., Tzeng, S.H., Chiu, Y.T., Characterizing elastic properties of carbon nanotube/polyimide nanocomposites using multi-scale simulation. *Composites: Part B Engineering* 2010, 41, 106–115. doi:10.1016/j.compositesb.2009.06.003

Tsai, K.-H., Kim, K.-S., 1996. The micromechanics of fiber pull-out. *Journal of the Mechanics and Physics of Solids* 44(7), 1147–1177. doi: 10.1016/0022-5096(96)00019-1

Tunvir, K., Kim, A., Nahm, S.H., 2008. The effect of two neighboring defects on the mechanical properties of carbon nanotubes. *Nanotechnology* 19, 065703. doi:10.1088/0957-4484/19/6/065703

Upadhyaya, P., Kumar, S., 2015. Micromechanics of stress transfer through the interphase in fiber-reinforced composites. *Mechanics of Materials* 89, 190–201. doi: 10.1016/j.mechmat.2015.06.012

Varshney, V., Patnaik, S.S., Roy, A.K., Farmer, B.L., 2008. A molecular dynamics study of epoxy-based networks: cross-linking procedure and prediction of molecular and material properties. *Macromolecules* 41(18)**,** 6837–6842. doi: 10.1021/ma801153e

Wang, Y., Weng, G.J., Meguid, S.A., Hamouda, A.M., 2014. A continuum model with a percolation threshold and tunneling-assisted interfacial conductivity for carbon nanotube-based nanocomposites. *Journal of Applied Physics*, 115, 193706. doi: 10.1063/1.4878195

Wang, Y., Shan, J.W., Weng, G.J., A.M., 2015. Percolation threshold and electrical conductivity of graphene-based nanocomposites with filler agglomeration and interfacial tunneling. *Journal of Applied Physics*, 118, 065101. doi: 10.1063/1.4928293

Wicks, S.S., Wang, W., Williams, M.R., Wardle, B.L., 2014. Multi-scale interlaminar fracture mechanisms in woven composite laminates reinforced with aligned carbon nanotubes. *Composites Science and Technology* 100, 128–135. doi:10.1016/j.compscitech.2014.06.003

Yang, S., Yu, S., Kyoung, W., Han, D.-S., Cho, M., 2012. Multiscale modeling of size-dependent elastic properties of carbon nanotube/polymer nanocomposites with interfacial imperfections. *Polymer* 53, 623–633. doi: 10.1016/j.polymer.2011.11.052

Zhang, J., Zhuang, R., Liu, J., Mäder, E., 2010. Heinrich G, Gao S. Functional interphases with multi-walled carbon nanotubes in glass fibre/epoxy composites. *Carbon* 48, 2273–2281. doi: 10.1016/j.carbon.2010.03.001

Zhang, X., Fan, X., Yan, C., Li, H., Zhu, Y., Li, X., Yu, L., 2012. Interfacial microstructure and properties of carbon fiber composites modified with graphene oxide. *ACS Applied Material Interfaces* 4, 1543–1552. doi:10.1021/ am201757v

## Appendix A: Elements of Eshelby, P, and Q tensors, and transformation matrix

In case of CNS are aligned along the direction of a microscale fiber, the elements of the Eshelby tensor $[S^E]$ are explicitly written as follows (Qiu and Weng, 1990):

$$S^E_{2222} = S^E_{3333} = \frac{5-4\nu^m}{8(1-\nu^m)},$$

$$S^E_{2211} = S^E_{3311} = \frac{\nu^m}{2(1-\nu^m)},$$

$$S^E_{2233} = S^E_{3322} = \frac{4\nu^m-1}{8(1-\nu^m)},$$

$$S^E_{1313} = S^E_{1212} = 1/4,$$

$$S_{2323} = \frac{3-4\nu^m}{8(1-\nu^m)} \text{ with all other elements being zero} \quad \text{(A1)}$$

Similarly, in case of CNS are aligned radially to the axis a microscale fiber, the elements of the Eshelby tensor $[S^E]$ can be written as follows:

$$S^E_{1111} = S^E_{2222} = \frac{5-4\nu^m}{8(1-\nu^m)},$$

$$S^E_{1122} = S^E_{2211} = \frac{4\nu^m-1}{8(1-\nu^i)},$$

$$S^E_{1133} = S^E_{2233} = \frac{\nu^m}{2(1-\nu^m)},$$

$$S^E_{1313} = S^E_{2323} = 1/4,$$

$$S^E_{1212} = \frac{3-4\nu^m}{8(1-\nu^m)} \text{ with all other elements being zero} \quad \text{(A2)}$$

Direction cosines corresponding to the transformation matrix [Eq. (5)]:

$$t_{11} = \cos\phi \cos\psi - \sin\phi \cos\gamma \sin\psi,$$

$$t_{12} = \sin\phi \cos\psi + \cos\phi \cos\gamma \sin\psi,$$

$$t_{13} = \sin\psi \sin\gamma\,,$$

$$t_{21} = -\cos\phi \sin\psi - \sin\phi \cos\gamma \cos\psi,$$

$$t_{22} = -\sin\phi \sin\psi + \cos\phi \cos\gamma \cos\psi,$$

$$t_{23} = \sin\gamma \cos\psi,$$

$$t_{31} = \sin\phi \sin\gamma,$$

$$t_{32} = -\cos\phi \sin\gamma \quad \text{and} \quad t_{33} = \cos\gamma \tag{A3}$$

In case CNS are aligned along the direction of a microscale fiber with imperfect bonding with the surrounding epoxy, the elements of P and Q tensors are explicitly written as follows (Barai and Weng, 2011):

$$P_{2222} = P_{3333} = 4P_{3131} = 4P_{2121} = 2P_{2323} = \frac{2\pi}{8a},$$

$$Q_{2222} = Q_{3333} = 3Q_{2233} = 3Q_{3322} = 3Q_{2323} = \frac{9\pi}{32a} \text{ with all other elements are zero (A4)}$$

where a is the equivalent radius of CNS.

## Appendix B: Explicit forms of constant coefficients

The constants ($A_i$) obtained in the course of deriving the shear lag model in section 3 are explicitly expressed as follows:

$$A_1 = \frac{1}{C_{66}^i}\left[a\ln\frac{r}{a} + \frac{a}{b^2 - a^2}\left\{a^2\ln\frac{r}{a} - \frac{r^2 - a^2}{2}\right\}\right], \qquad A_2 = -\frac{b}{C_{66}^i(b^2 - a^2)}\left[a^2\ln\frac{r}{a} - \frac{r^2 - a^2}{2}\right],$$

$$A_3 = \frac{a\ln(b/a)}{C_{66}^i} + \frac{a}{C_{66}^i(b^2 - a^2)}\left\{a^2\ln\frac{b}{a} - \frac{b^2 - a^2}{2}\right\},$$

$$A_4 = \frac{b^3 - a^2 b}{2C_{66}^i(c^2 - b^2)} + \frac{b}{C_{66}^i(b^2 - a^2)}\left\{a^2\ln\frac{a}{b} - \frac{a^2 - b^2}{2}\right\},$$

$$A_5 = \frac{C_{13}^{i}}{C_{66}^{i}}\left[a\ln(b/a) + \frac{a}{b^2 - a^2}\left\{a^2\ln\frac{b}{a} - \frac{b^2 - a^2}{2}\right\}\right], \qquad A_6 = A_2|_{r=b,} \quad A_7 = A_3|_{r=b,}$$

$$A_8 = A_4|_{r=b,} \quad A_9 = A_4|_{r=c,} \quad , \qquad A_{10} = \frac{2b^2C_{12}^{i}}{b^2 - a^2}, \qquad A_{11} = \frac{2C_{12}^{i}}{b^2 - a^2}$$

$$A_{12} = \frac{aC_{11}^{i}}{C_{66}^{i}(b^2 - a^2)}\left[b^2\ln\frac{b}{a} - \frac{b^2 - a^2}{2} + \frac{a^2b^2\ln(b/a)}{2(b^2 - a^2)} - \frac{b^4 - a^4}{8(b^2 - a^2)}\right],$$

$$A_{13} = -\frac{a^2b^3C_{11}^{i}\ln(b/a)}{C_{66}^{i}(b^2 - a^2)^2}, \qquad A_{14} = \frac{C_{11}^{i}}{C_{11}^{f}} + A_9b_{11} + A_{10}b_{21} + A_{11}b_{31},$$

$$A_{15} = A_9b_{12} + A_{10}b_{22} + A_{11}b_{32} + A_{12}, \qquad A_{16} = A_9b_{13} + A_{10}b_{23} + A_{11}b_{33} + A_{13},$$

$$A_{17} = \frac{C_{11}^{m}}{C_{11}^{f}} - \frac{2C_{11}^{m}C_{12}^{f}}{C_{11}^{f}}b_{11} + 2C_{12}^{m}b_{21}, \qquad A_{18} = -\frac{2C_{11}^{m}C_{12}^{f}}{C_{11}^{f}}b_{12} + 2C_{12}^{m}b_{22} + C_{11}^{m}A_3,$$

$$A_{19} = -\frac{2C_{11}^{m}C_{12}^{f}}{C_{11}^{f}}b_{13} + 2C_{12}^{m}b_{23} + C_{11}^{m}A_4, \qquad A_{20} = a^2 + (b^2 - a^2)A_{14} + (c^2 - b^2)A_{17},$$

$$A_{21} = -\frac{a}{2}[(b^2 - a^2)A_{15} + (c^2 - b^2)A_{18}], \qquad A_{22} = \frac{c^2 - b^2}{2b}[(b^2 - a^2)A_{16} + (c^2 - b^2)A_{19}],$$

$$A_{23} = -\frac{a}{2}\left[\frac{A_{15}A_{19}}{A_{16}} - A_{18}\right],$$

$$A_{24} = \frac{1}{A_{23}}\left[\frac{A_{14}A_{19}}{A_{16}} - A_{17} + \frac{a^2A_{19}}{(b^2 - a^2)A_{16}} - \frac{A_{21}A_{19}(c^2 - b^2)}{A_{16}A_{22}(b^2 - a^2)} - \frac{A_{21}}{A_{22}}\right],$$

$$A_{25} = -\frac{1}{A_{23}}\left[\frac{A_{20}}{A_{22}} + \frac{A_{19}A_{20}(c^2 - b^2)}{A_{16}A_{22}(b^2 - a^2)}\right], \qquad A_{26} = -\frac{1}{A_{23}}\left[\frac{a^2}{A_{22}} + \frac{A_{19}a^2(c^2 - b^2)}{A_{16}A_{22}(b^2 - a^2)}\right],$$

$$A_{31} = \frac{a}{2}\frac{A_{18}A_{20}}{A_{17}} + A_{21}, \qquad A_{32} = \left(\frac{c^2 - b^2}{2b}\right)\frac{A_{19}A_{20}}{A_{17}} - A_{22}, \qquad A_{33} = A_{14} - \frac{A_{15}A_{17}}{A_{18}},$$

$$A_{34} = A_{16} - \frac{A_{15}A_{19}}{A_{18}}, \qquad A_{35} = A_{34}(b^2 - a^2)\left(\frac{c^2 - b^2}{2b}\right),$$

$$A_{36} = \frac{1}{A_{35}}\left[\frac{a^2 A_{32}}{A_{31}} + (b^2 - a^2)\left(\frac{A_{15}}{A_{18}} + \frac{A_{32}A_{33}}{A_{31}}\right) + (c^2 - b^2)\right],$$

$$A_{37} = -\frac{1}{A_{35}}\left[\frac{a^2 A_{20}}{A_{17}A_{31}} + \frac{A_{20}A_{33}}{A_{17}A_{31}}(b^2 - a^2)\right] \text{ and } A_{38} = -\frac{1}{A_{35}}\left[\frac{a^4}{A_{31}} + a^2(b^2 - a^2)\frac{A_{33}}{A_{31}}\right] \quad \text{(B1)}$$

The constants $b_{ij}$ appeared in Eq. (33) are expressed as follows:

$$B_{11} = C_{23}^{i} - C_{23}^{f} - C_{33}^{f} - C_{33}^{i}\frac{a^2 + b^2}{b^2 - a^2} - \frac{2C_{12}^{f}C_{13}^{i}}{C_{11}^{f}} + \frac{2(C_{12}^{f})^2}{C_{11}^{f}}, \qquad B_{12} = \frac{2C_{33}^{i}b^2}{b^2 - a^2},$$

$$B_{13} = \frac{2C_{33}^{i}}{b^2 - a^2}, \qquad B_{21} = \frac{2C_{33}^{i}a^2}{b^2 - a^2} + \frac{2C_{13}^{i}C_{12}^{f}}{C_{11}^{f}} - \frac{2C_{12}^{m}C_{12}^{f}}{C_{11}^{f}}$$

$$B_{22} = C_{11}^{f} + C_{12}^{m} - C_{23}^{i} - C_{33}^{i}\frac{a^2 + b^2}{b^2 - a^2}, \qquad B_{23} = \frac{1}{b^2}\left(C_{12}^{m} - C_{11}^{m} - C_{23}^{i} - C_{33}^{i}\frac{a^2 + a^2}{b^2 - a^2}\right)$$

$$B_{31} = -\frac{2C_{12}^{f}C_{12}^{m}}{C_{11}^{f}}, \qquad B_{32} = C_{11}^{f} + C_{12}^{m} \text{ and } B_{33} = \frac{1}{c^2}(C_{12}^{m} - C_{11}^{m}) \qquad \text{(B2)}$$

**Table 1** Parameters used for MD simulations

| **Parameter** | **Epoxy** | **Bundle of 13 CNTs** |
|---|---|---|
| CNT type | - | (5, 5) |
| Number of CNTs | - | 13 |
| Length of a CNT (Å) | - | 100 |
| CNT volume fraction | - | ~14% |
| RVE dimensions ($Å^3$) | 39×39×39 | 75×75×100 |
| Total number of EPON 862 molecules | 100 | 800 |
| Total number of DETDA molecules | 50 | 400 |
| Total number of atoms | 6250 | 61050 |

**Table 2** Elastic moduli of the epoxy material

| | Young's modulus (GPa) | Poisson's ratio |
|---|---|---|
| Present MD simulations | 3.5 | 0.36 |
| Experimental (Morris, 2008) | 3.43 | - |

**Table 3** Effective elastic coefficients of the CNS and corresponding displacement fields

| **Elastic coefficients** | **Applied strains** | **Applied displacement** |
|---|---|---|
| $C_{11}$ | $\varepsilon_{11} = e$ | $u_1 = ex_1$ |
| $C_{22} = C_{33}$ | $\varepsilon_{22} = e$ | $u_2 = ex_2$ |
| $C_{44}$ | $\varepsilon_{23} = e/2$ | $u_2 = \frac{e}{2}x_3,$ $u_3 = \frac{e}{2}x_2$ |
| $C_{55} = C_{66}$ | $\varepsilon_{13} = e/2$ | $u_1 = \frac{e}{2}x_3,$ $u_3 = \frac{e}{2}x_1$ |
| $K_{23} = \frac{C_{22} + C_{23}}{2}$ | $\varepsilon_{22} = \varepsilon_{33}$ | $u_2 = ex_2,$ $u_3 = ex_3$ |

**Table 4** Effective elastic properties of the CNS and the interphase with perfectly bonded CNS-epoxy interfaces

| **RVE** | **Modeling Technique** | **CNT Volume Fraction (%)** | $C_{11}$ **(GPa)** | $C_{12}$ **(GPa)** | $C_{13}$ **(GPa)** | $C_{33}$ **(GPa)** | $C_{44}$ **(GPa)** | $C_{66}$ **(GPa)** |
|---|---|---|---|---|---|---|---|---|
| CNS | MD simulations | 19% loading in the CNS | 108.35 | 3.5 | 3.5 | 6.7 | 1.48 | 1.83 |
| Interphase containing aligned CNS | MD simulations in conjunction with Mori-Tanaka model | 15% loading in the interphase | 86.93 | 3.46 | 3.46 | 6.51 | 1.39 | 1.44 |
| Interphase containing aligned CNS | MD simulations in conjunction with Mori-Tanaka model | 10% loading in the interphase | 60.21 | 3.4 | 3.4 | 6.3 | 1.39 | 1.56 |
| Interphase containing aligned CNS | MD simulations in conjunction with Mori-Tanaka model | 5% loading in the interphase | 32.92 | 3.36 | 3.36 | 6.08 | 1.33 | 1.42 |
| Interphase containing CNS, which are radially aligned to the axis of a microscale fiber | MD simulations in conjunction with Mori-Tanaka model | 5% loading in the interphase | 6.08 | 3.4 | 3.36 | 32.92 | 1.42 | 1.33 |
| Interphase containing randomly dispersed CNS | MD simulations in conjunction with Mori-Tanaka model and then transforming the properties | 5% loading in the interphase | 10.3 | 4.93 | 4.93 | 10.3 | 2.9 | 2.9 |

**Table 5** Effective elastic properties of the interphase with perfect and imperfect CNS-epoxy interfacial bonding conditions

| **RVE** | **α (nm/GPa)** | **CNT Volume Fraction (%)** | $C_{11}$ **(GPa)** | $C_{12}$ **(GPa)** | $C_{13}$ **(GPa)** | $C_{33}$ **(GPa)** | $C_{44}$ **(GPa)** | $C_{66}$ **(GPa)** |
|---|---|---|---|---|---|---|---|---|
| Interphase containing aligned CNS | 0 (perfectly bonded) | 5% loading in the interphase | 32.92 | 3.36 | 3.36 | 6.08 | 1.33 | 1.42 |
| Interphase containing aligned CNS | $0.5 \times 10^{-6}$ | 5% loading in the interphase | 32.92 | 3.34 | 3.34 | 3.36 | 1.32 | 1.4 |
| Interphase containing aligned CNS | $0.75 \times 10^{-6}$ | 5% loading in the interphase | 32.92 | 3.34 | 3.34 | 1.98 | 1.32 | 1.4 |

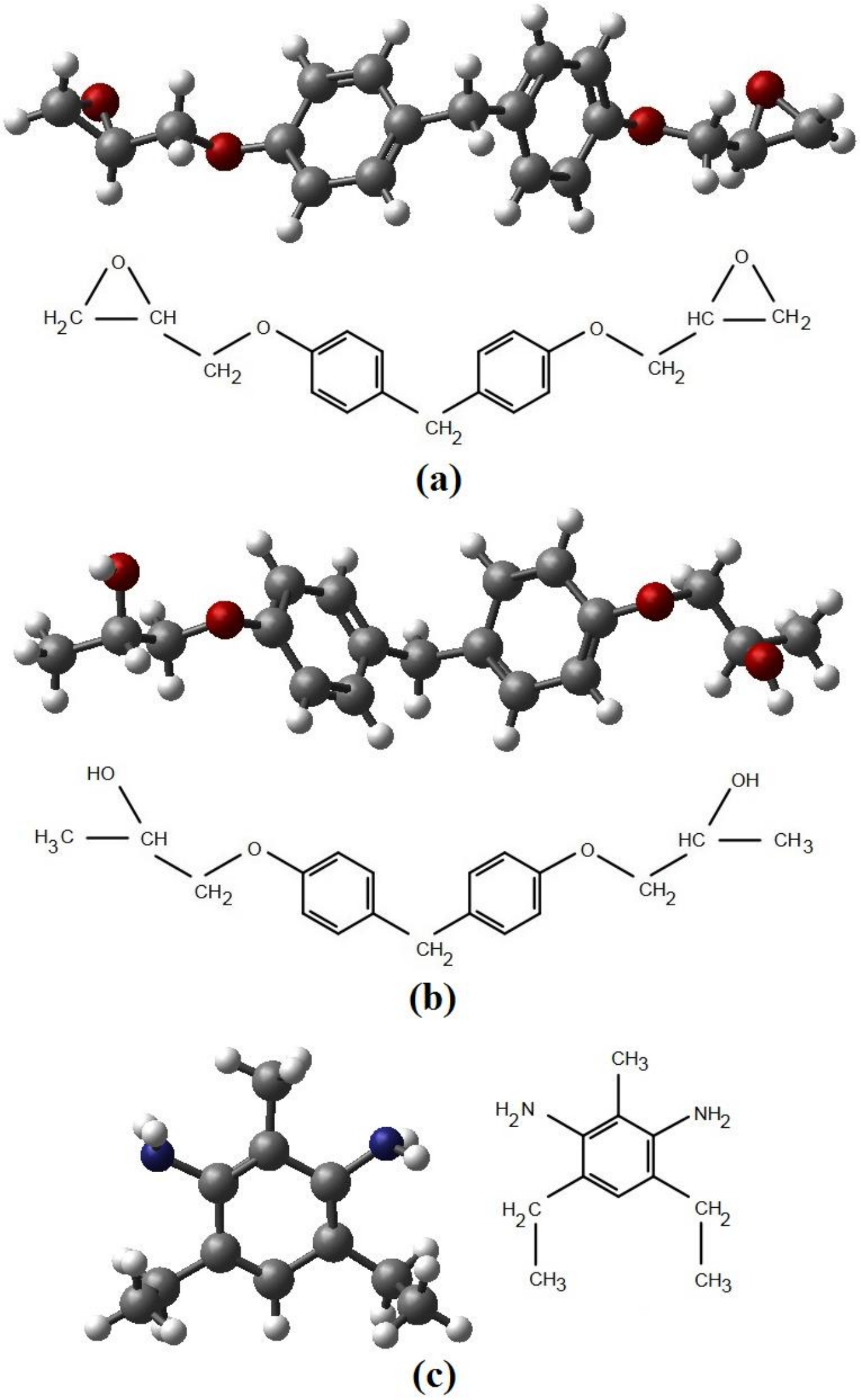

**Fig. 1.** Molecular and chemical structures of (a) EPON 862, (b) activated EPON 862 and (c) DETDA curing agent. Colored with gray, red, blue, and white are carbon, oxygen, nitrogen, and carbon atoms, respectively.

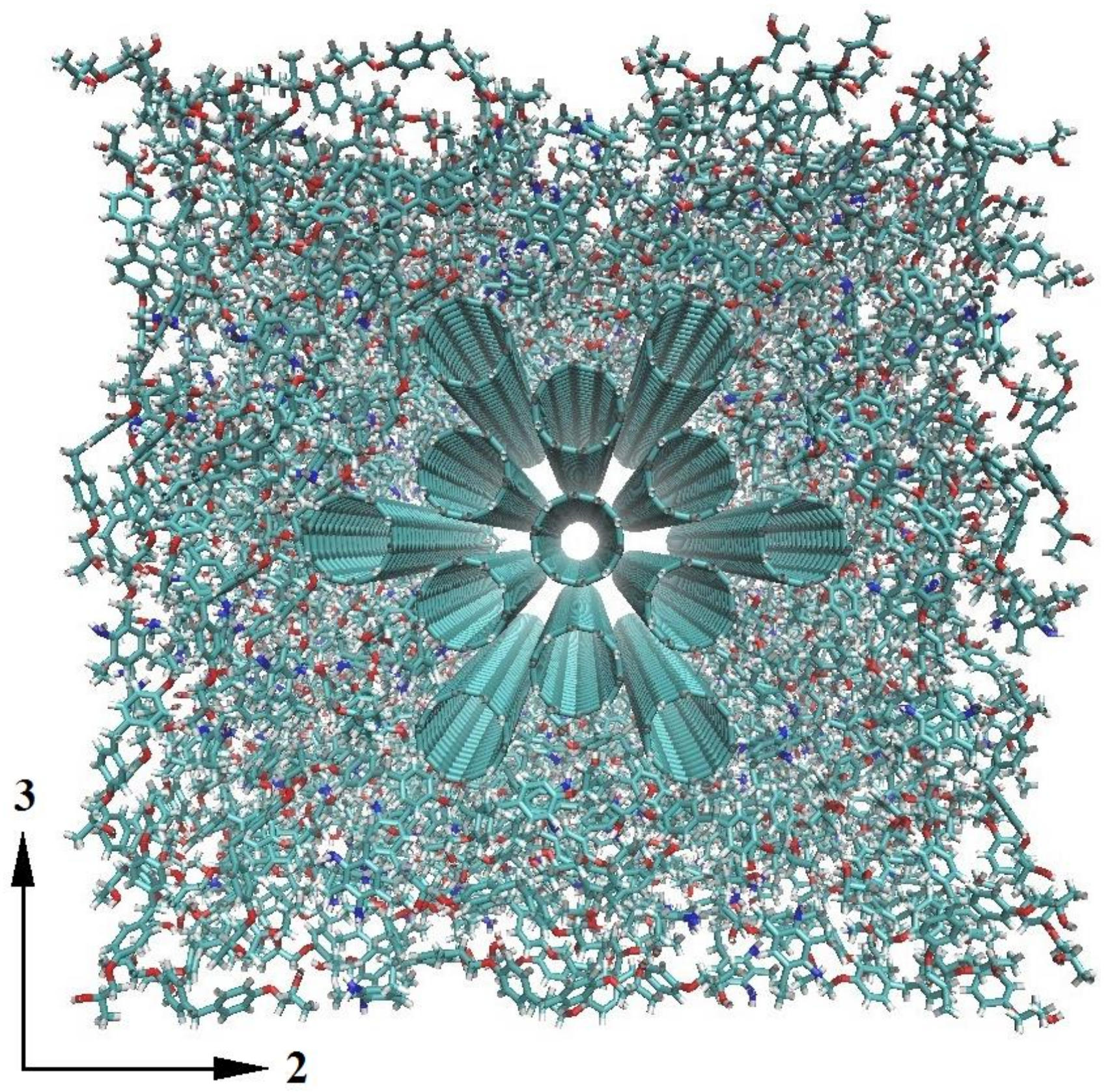

**Fig. 2.** MD unit cell, representing CNS, comprised of EPON 862-DETDA epoxy and a bundle of thirteen CNTs.

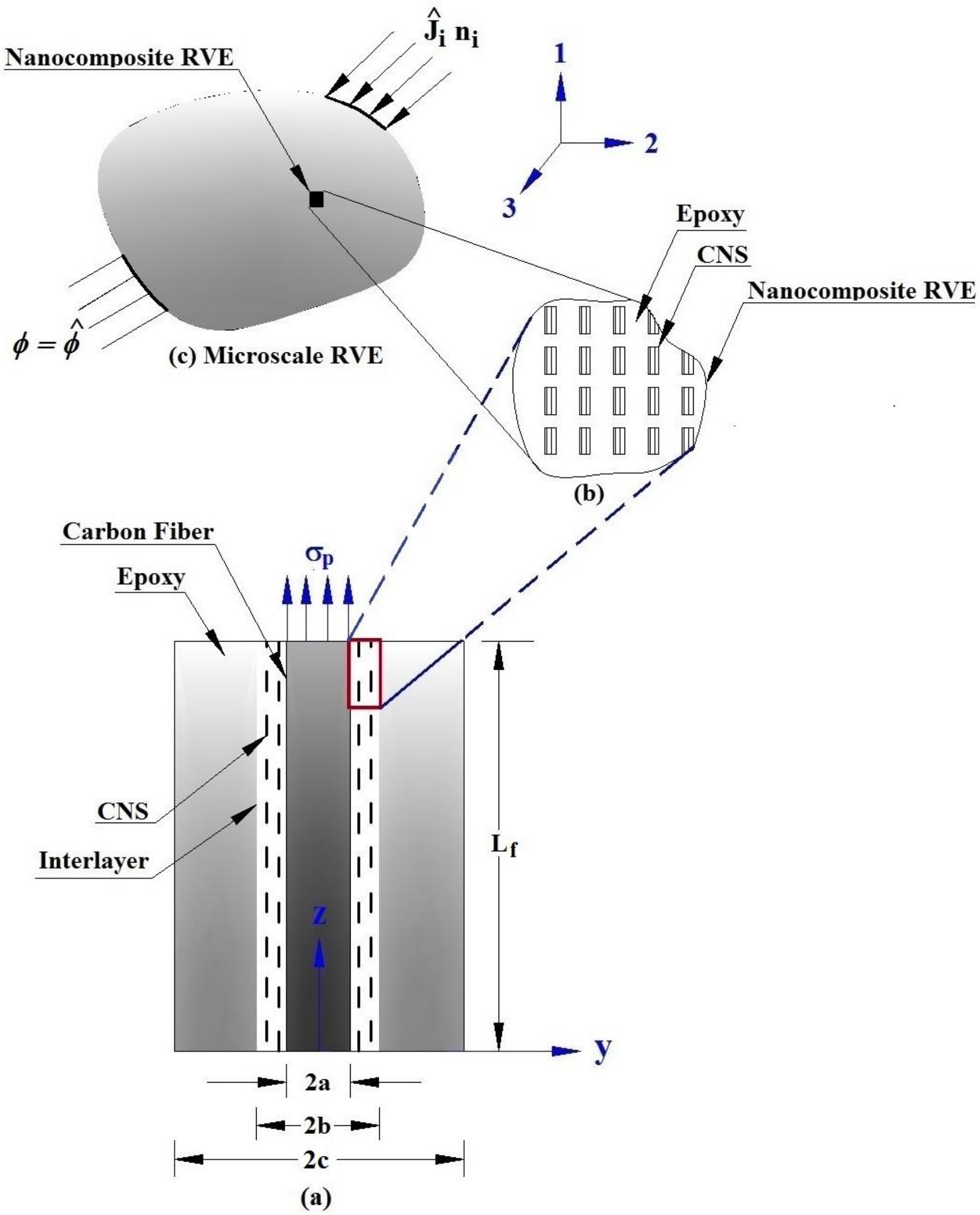

**Fig. 3.** Conceptual illustration of the multiscale composite: (a) three-phase pull-out model, (b) nanocomposite RVE comprised of CNS and epoxy, and (c) microscale RVE representing interphase.

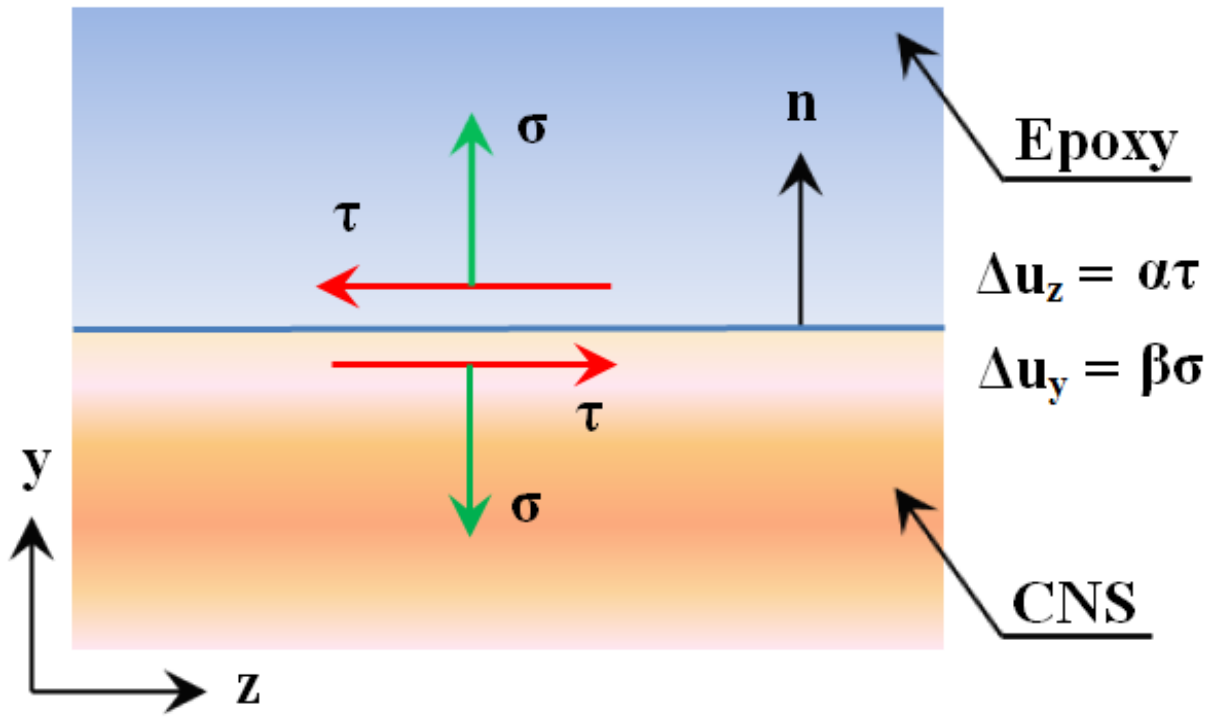

**Fig. 4.** CNS-epoxy interfacial conditions.

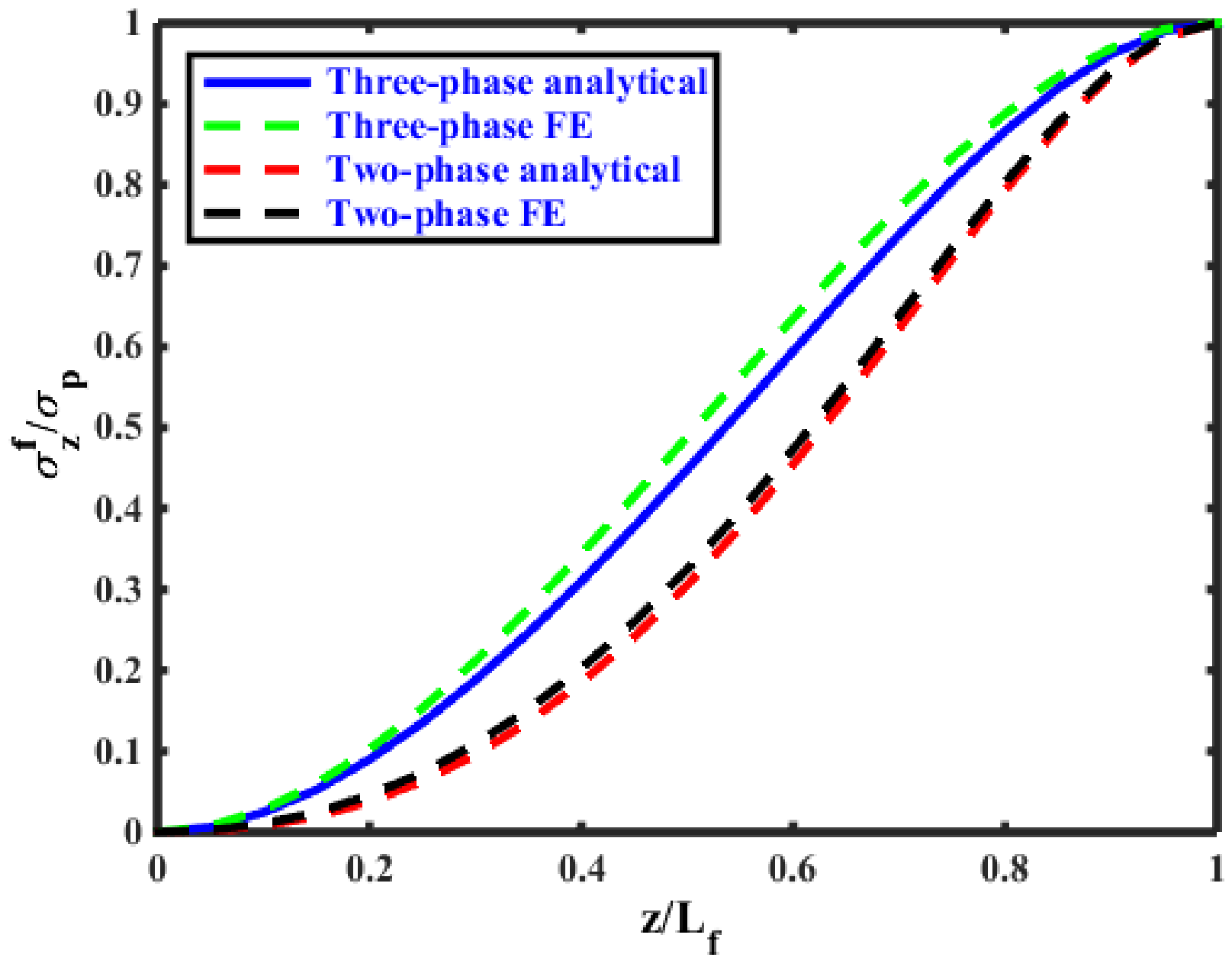

**Fig. 5.** Comparison of normalized axial stress in the fiber along its normalized length predicted by analytical model with those of FE simulations. For three-phase pull-out model: A-CNT bundles are parallel to the microscale fiber and their volume fraction in the interphase is 5%.

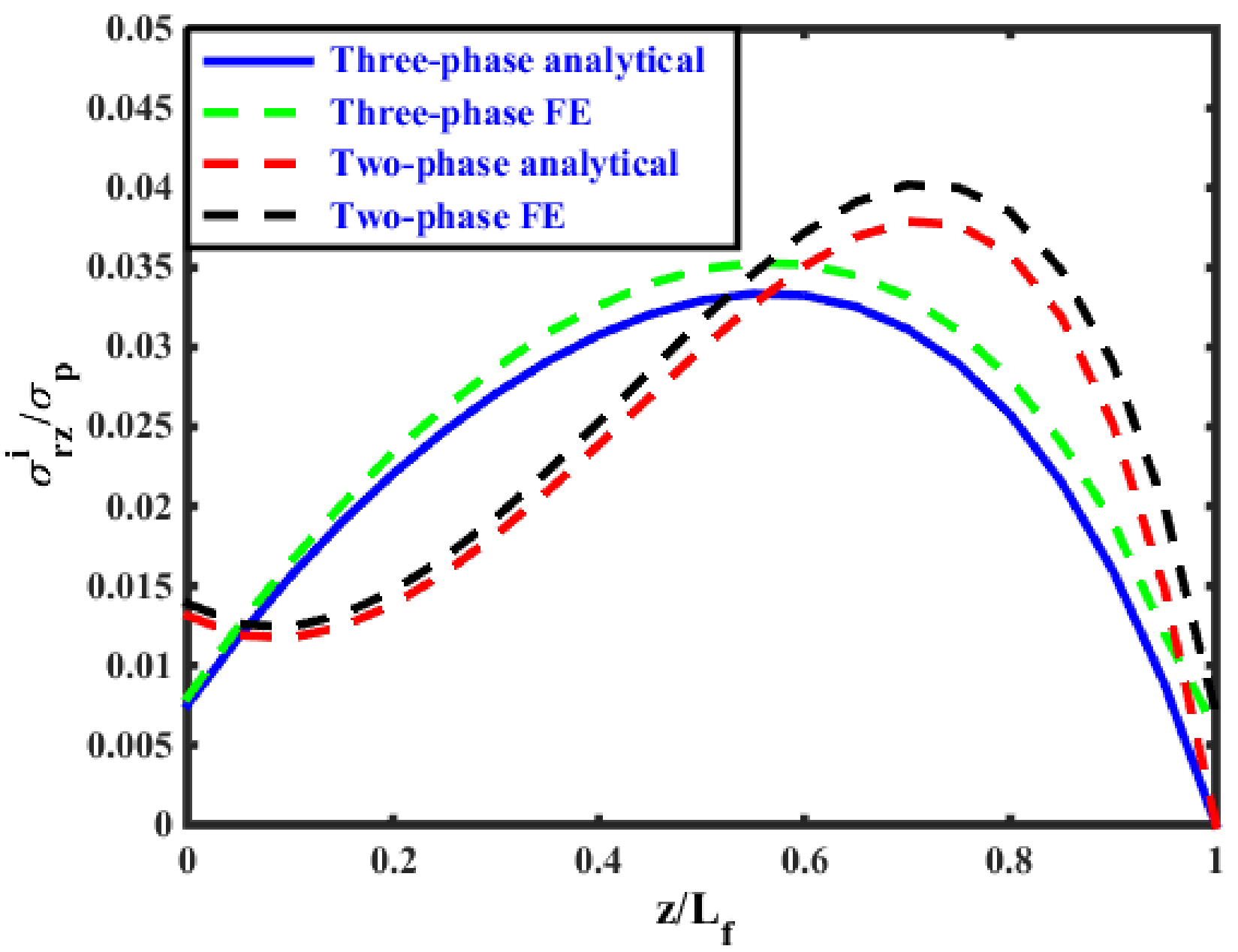

**Fig. 6.** Comparison of normalized transverse shear stress along the normalized length of the interphase, at r = 5.1 µm, predicted by analytical model with those of FE simulations. For three-phase pull-out model: A-CNT bundles are parallel to the microscale fiber and their volume fraction in the interphase is 5%.

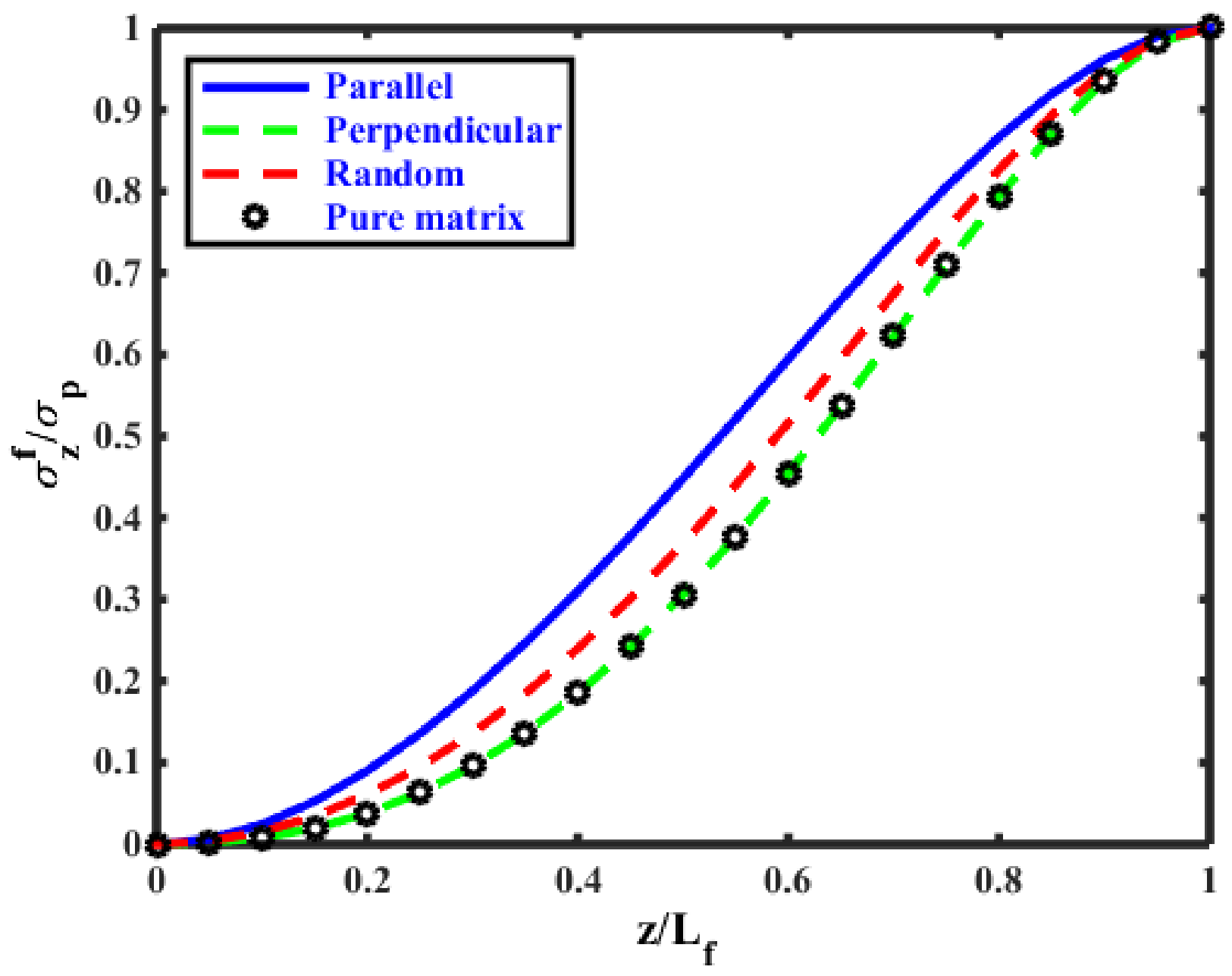

**Fig. 7.** Normalized axial stress in the fiber along its normalized length for different orientations of CNS (for 5% CNT loading in the interphase).

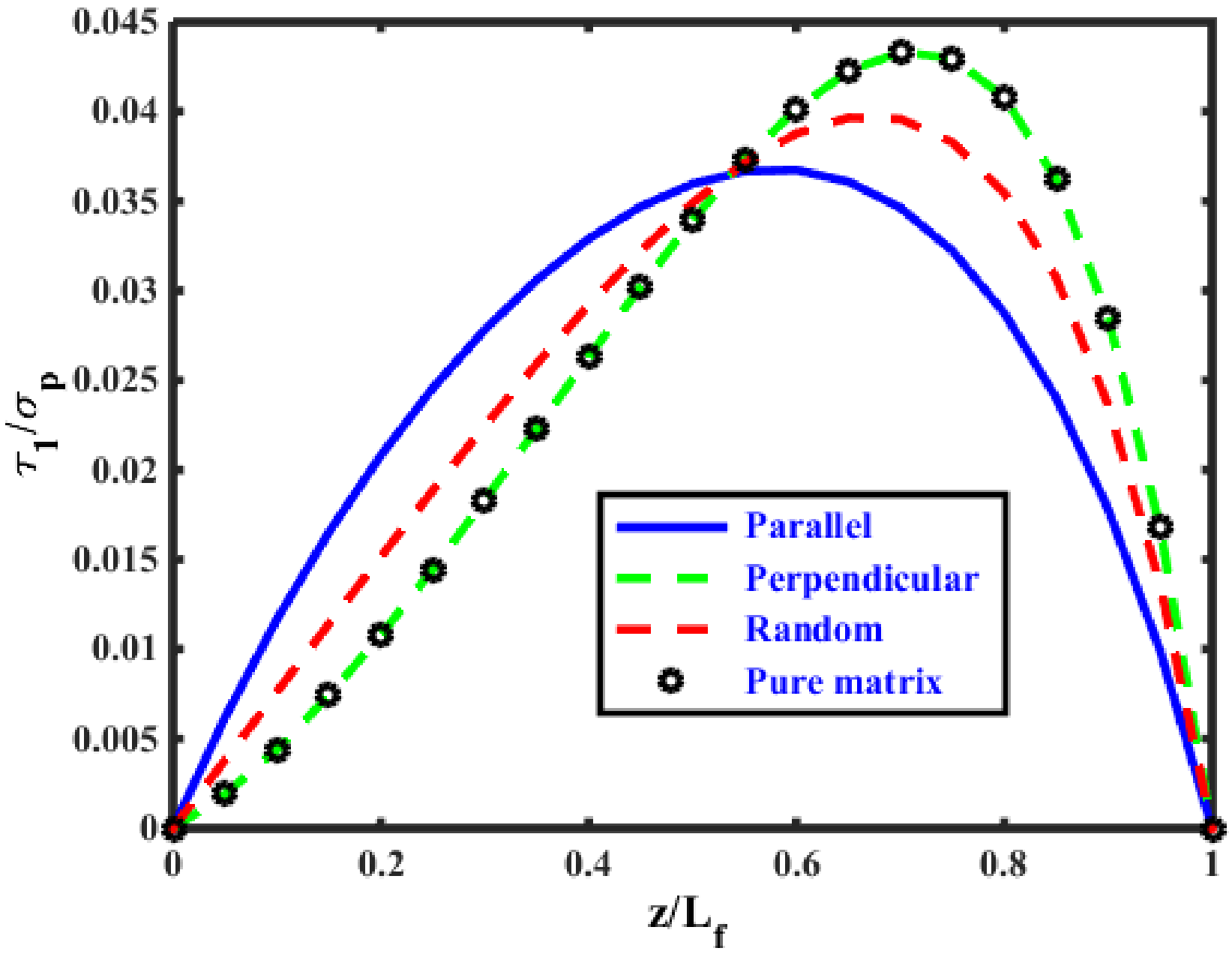

**Fig. 8.** Normalized interfacial shear stress at r=a along the normalized length of the fiber for different orientations of CNS (for 5% CNT loading in the interphase).

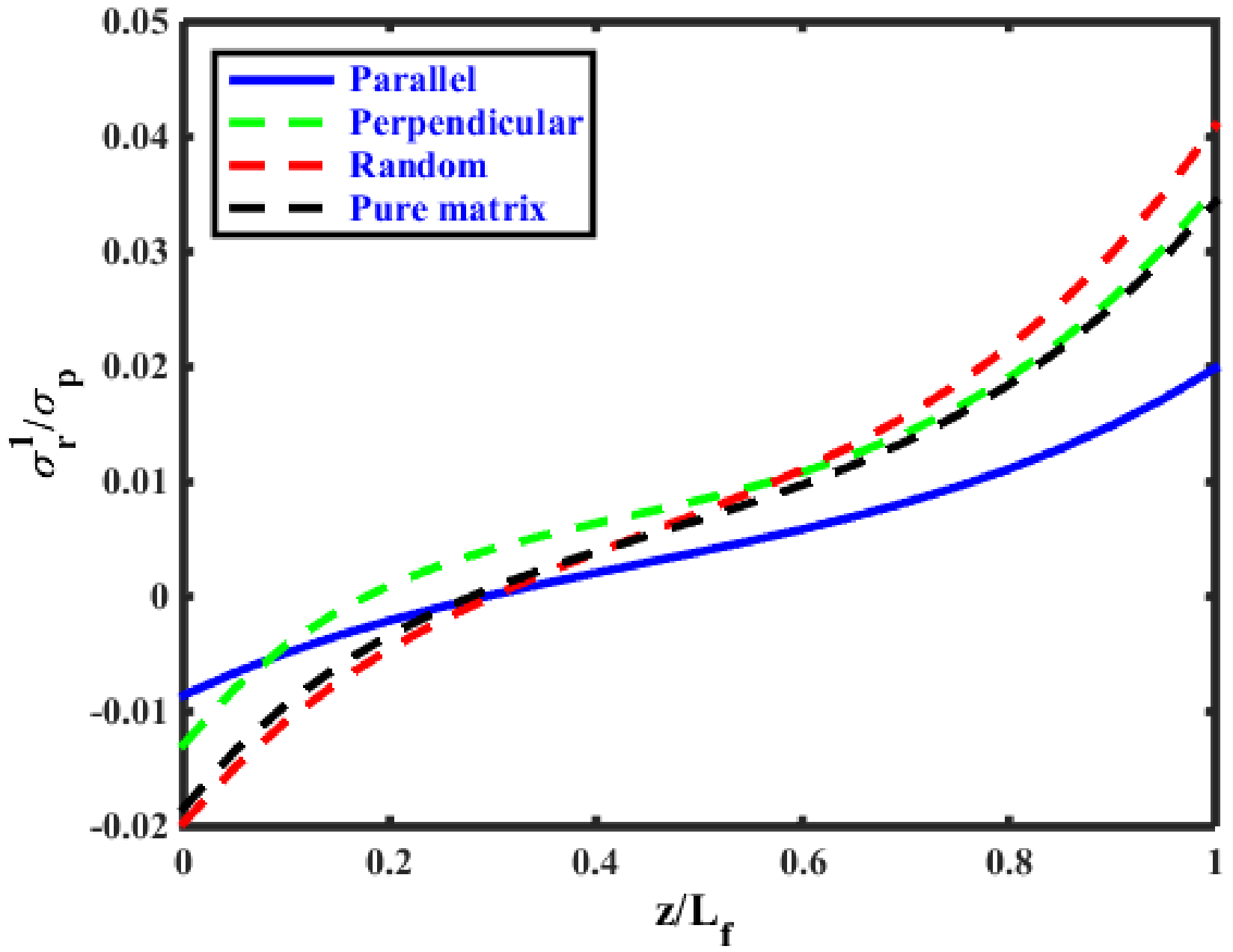

**Fig. 9.** Normalized interfacial radial stress at r = a over the normalized length of the microscale fiber for different orientations of CNS (for 5% CNT loading in the interphase).

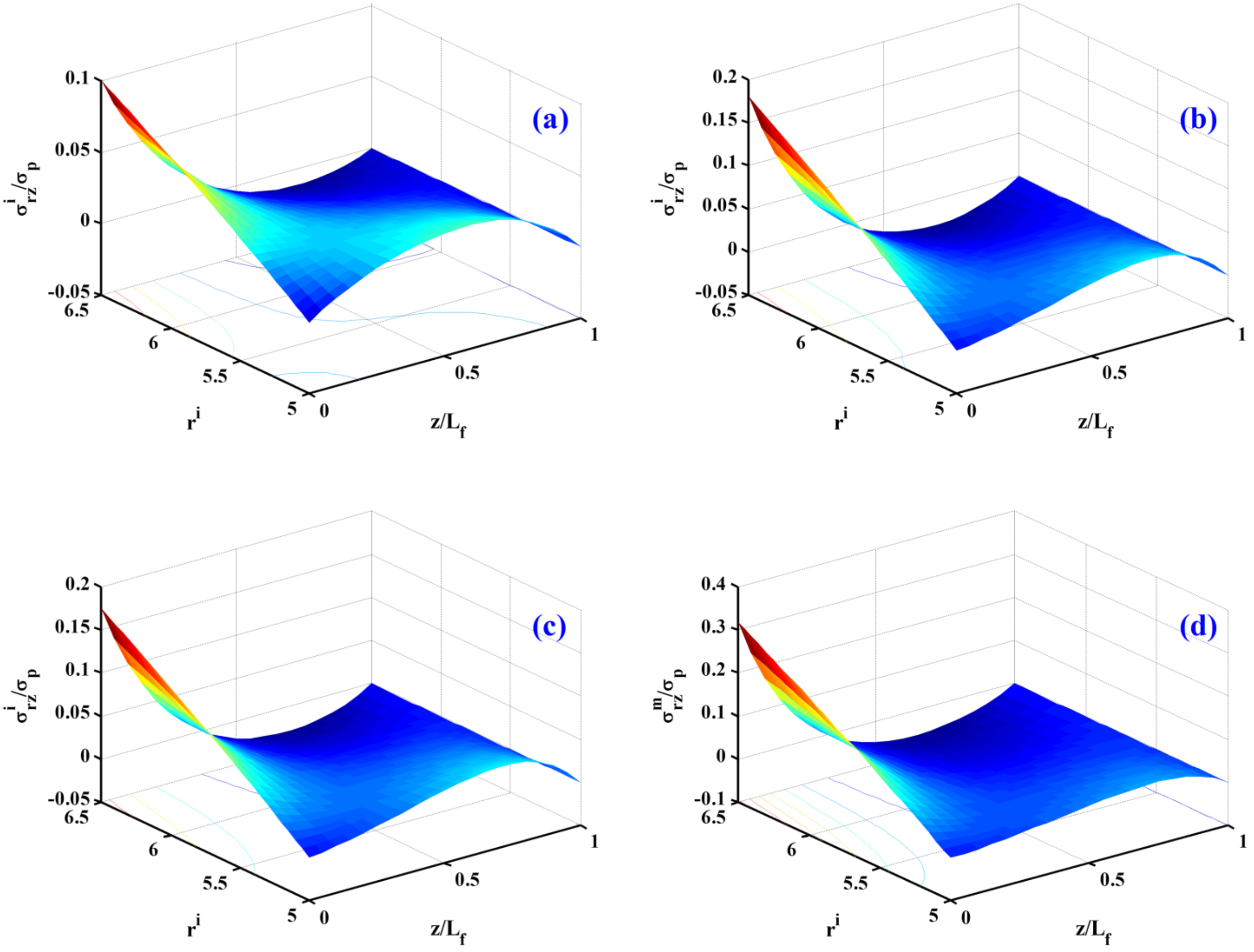

**Fig. 10.** Surface plots depicting the variation of transverse shear stress in the rz plane of interlayer and matrix over the entire length of the fiber: (a) parallel, (b) perpendicular, (c) random and (d) pure matrix.

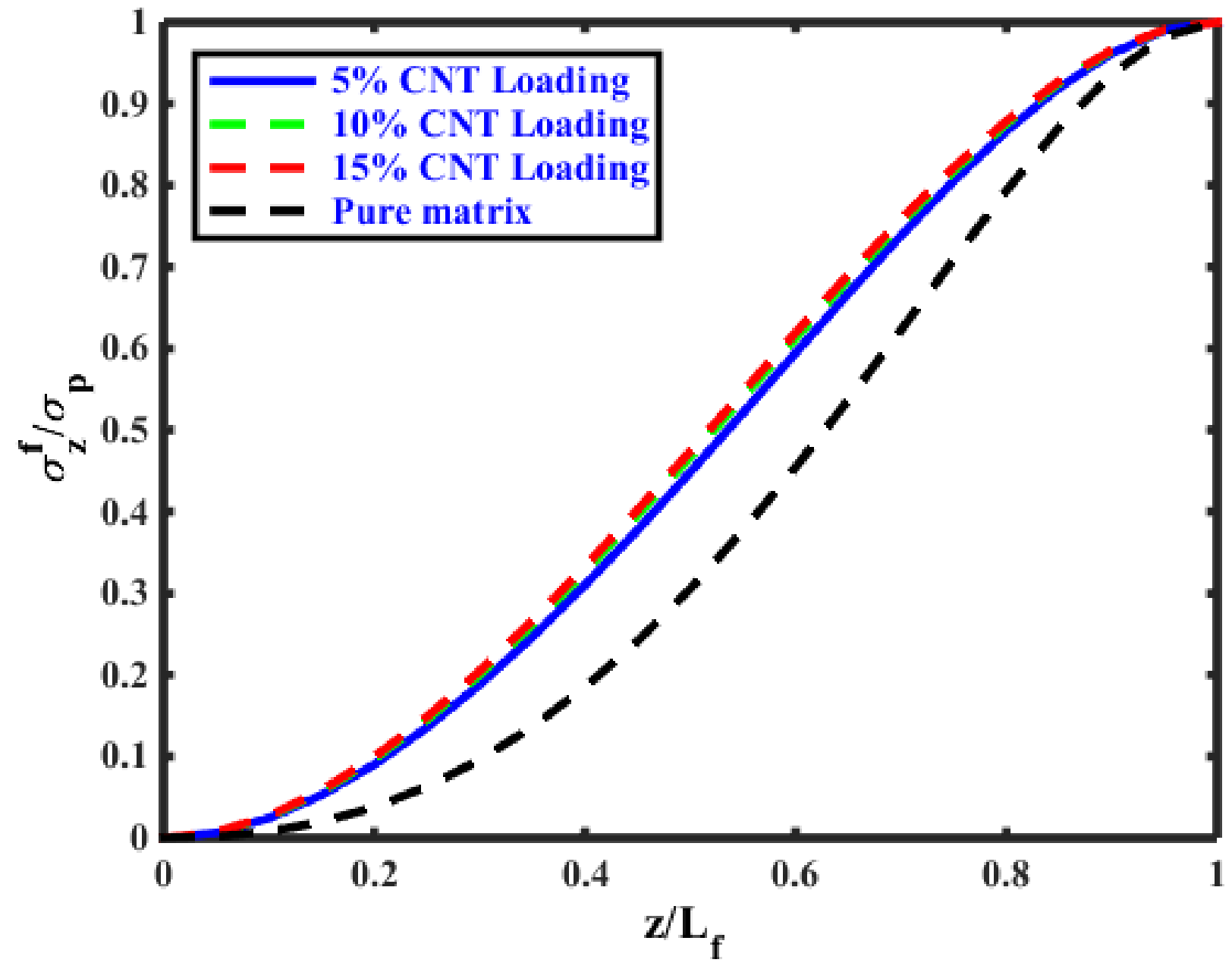

**Fig. 11.** Normalized axial stress in the microscale fiber along its length for different CNT loadings where CNS are parallel to the microscale fiber.

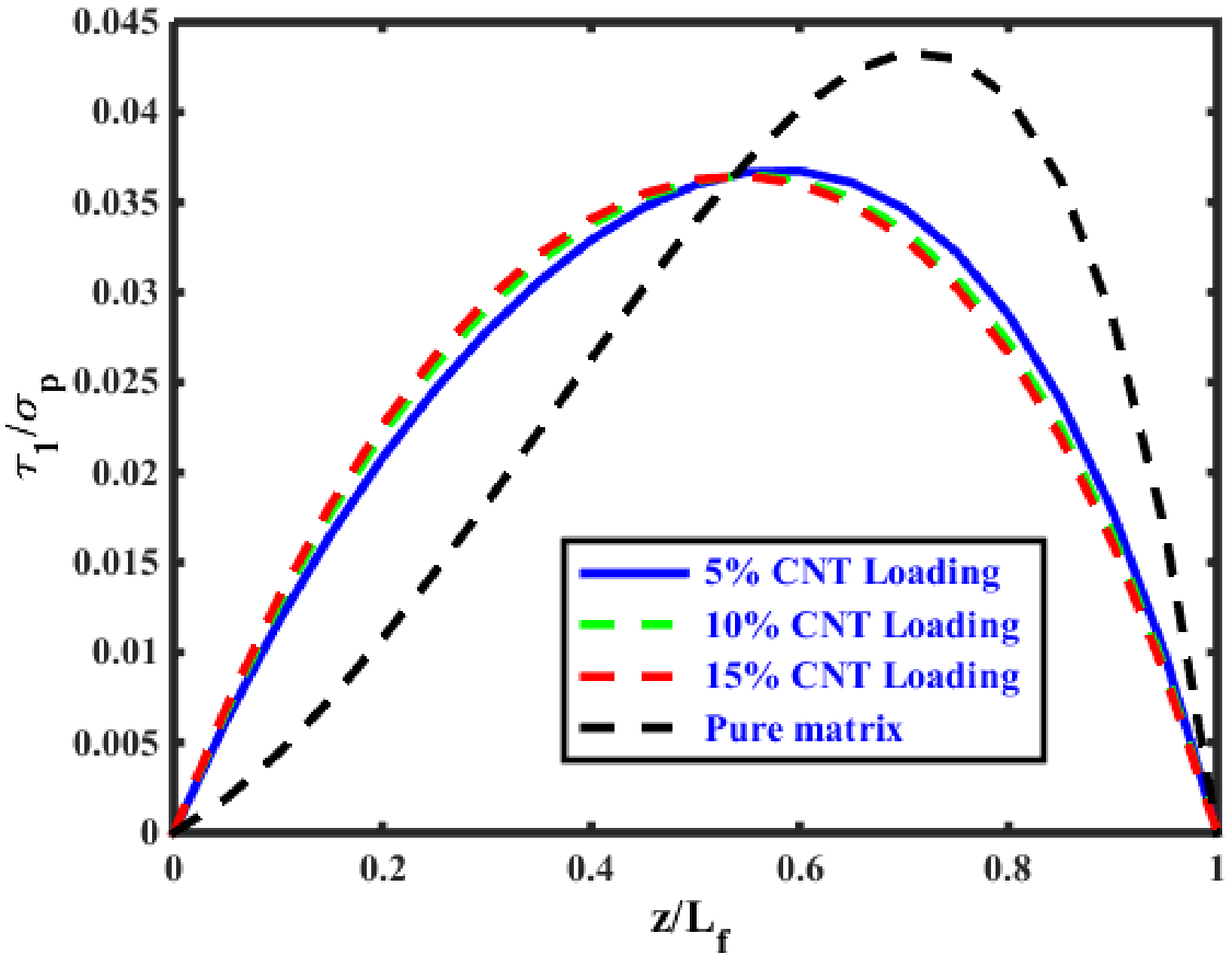

**Fig. 12.** Normalized interfacial shear stress at r=a along the length of the microscale fiber for different CNT loadings when CNS are parallel to the microscale fiber.

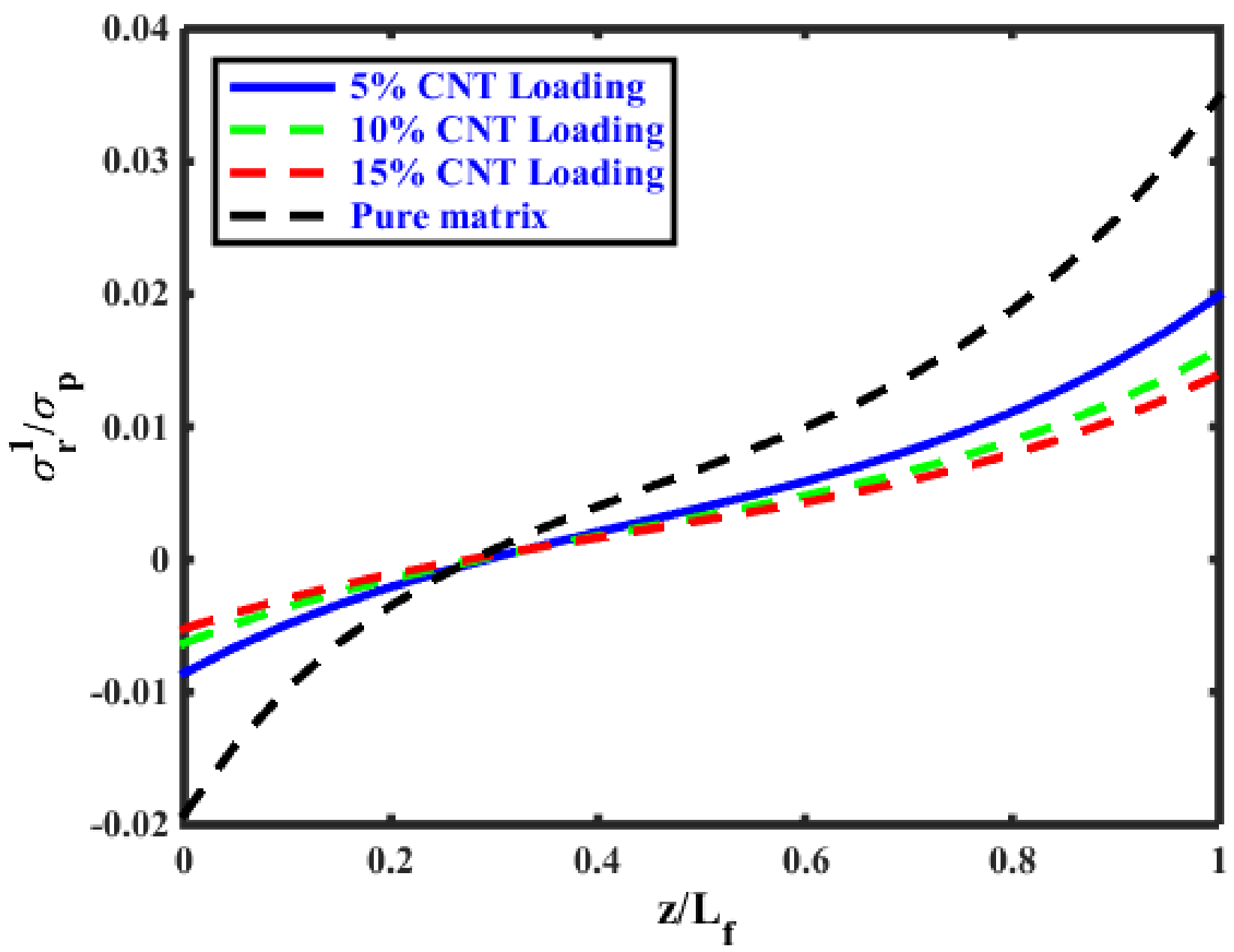

**Fig. 13.** Normalized interfacial radial stress at r = a along the length of the microscale fiber for different CNT loadings where CNS are parallel to the microscale fiber.

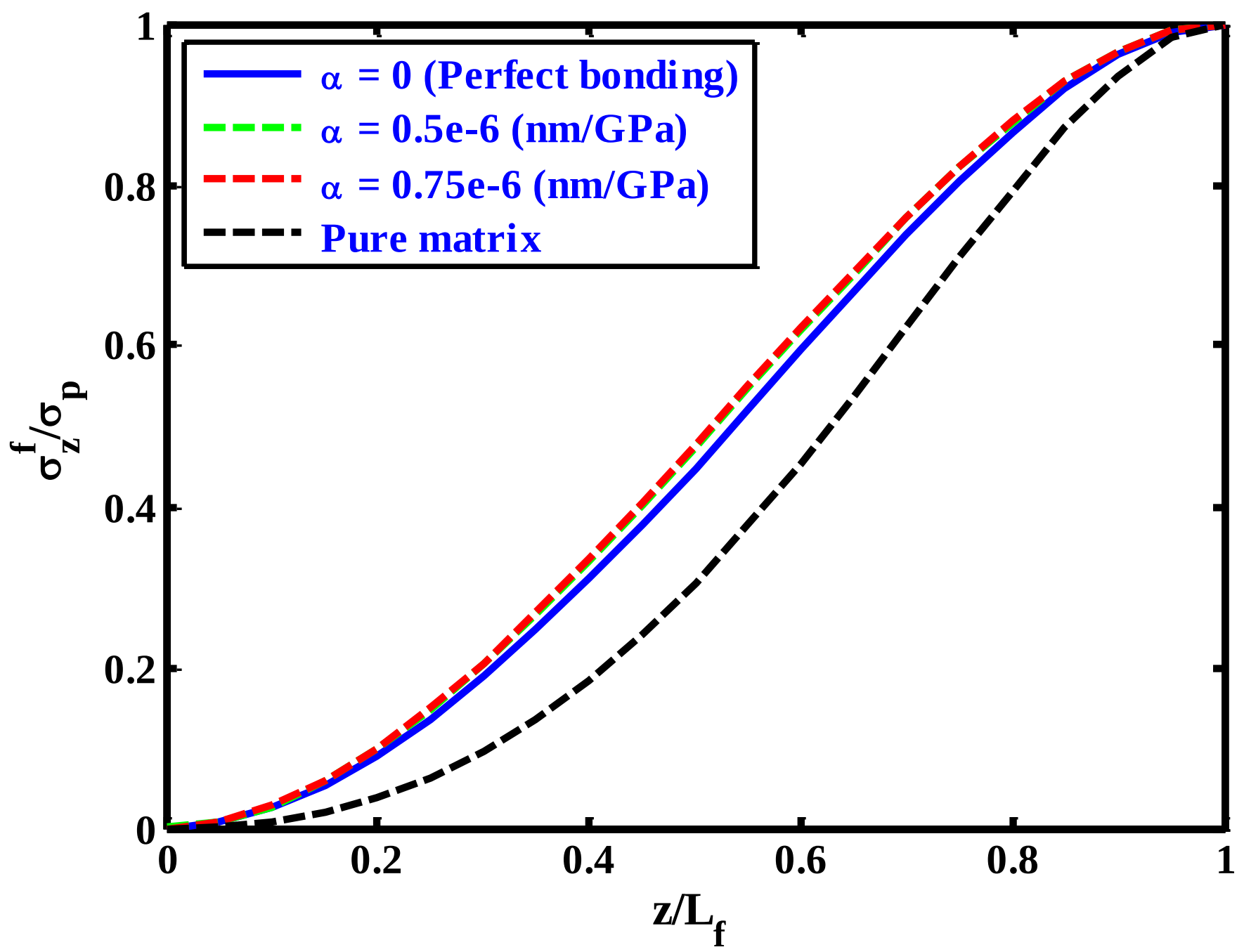

**Fig. 14.** Normalized axial stress in the microscale fiber along its length for different CNS-epoxy interfacial bonding conditions (CNS are parallel to the microscale fiber).

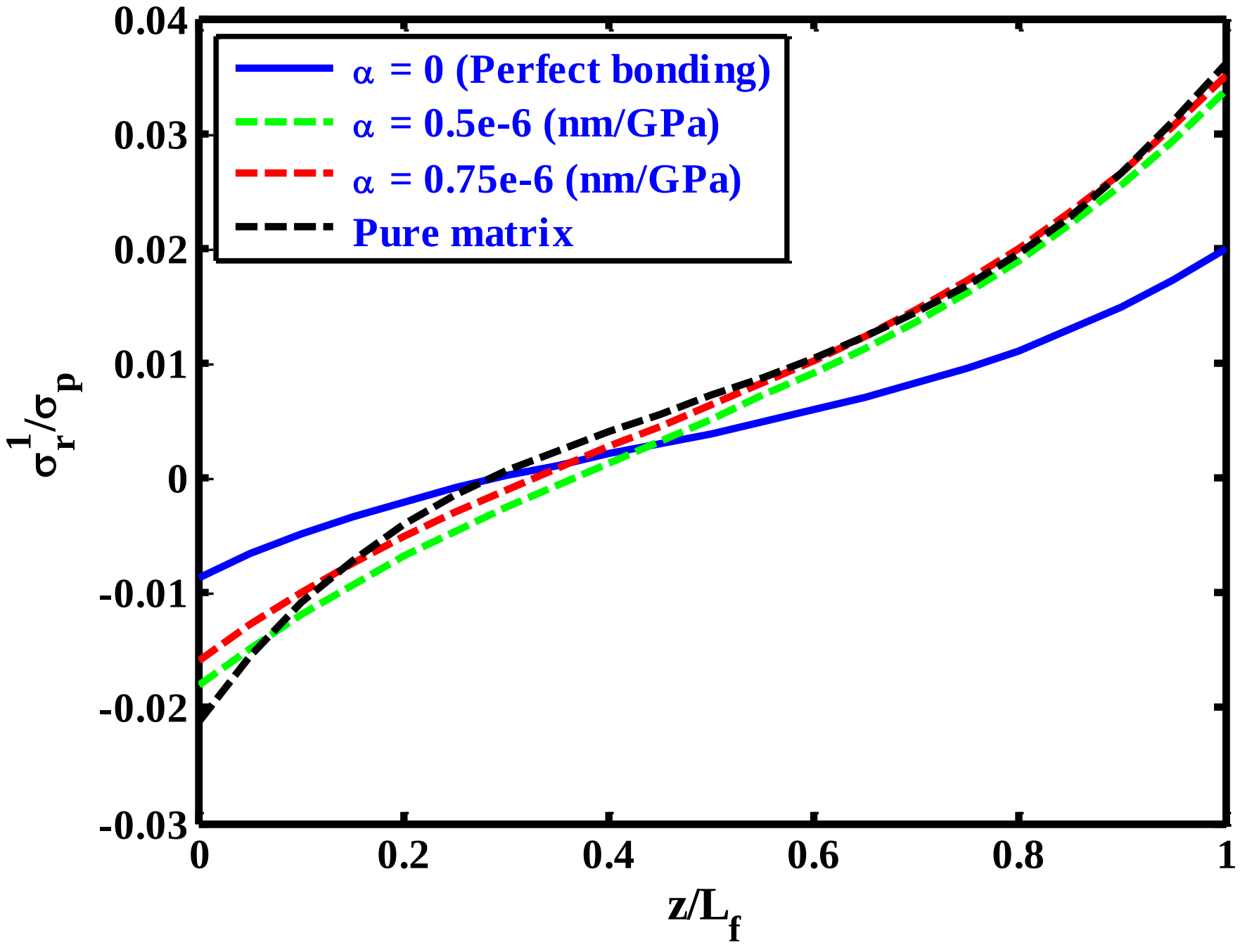

**Fig. 15.** Normalized interfacial radial stress at r = a along the length of the microscale fiber for different CNS-epoxy interfacial bonding conditions (CNS are parallel to the microscale fiber).